%% file: main.tex
\documentclass[preprint,10pt]{elsarticle}
\usepackage{url}

\usepackage{calc,xspace}
\usepackage[modulo]{lineno}  

\usepackage{amssymb} 
\usepackage{amsmath}
\usepackage[utf8]{inputenc}
\usepackage{amsbsy}
\usepackage{amsthm} 
\usepackage{amsfonts} 
\usepackage{euscript} 
\usepackage{graphicx} 
\usepackage{stmaryrd} 
\usepackage{mathrsfs}
\usepackage{color}

\include{macros}

\newtheorem{example}{Example}
\newtheorem{definition}{Definition}

\newtheorem{lemma}{Lemma}
\newtheorem{corollary}{Corollary}
\newtheorem{proposition}{Proposition}
\newtheorem{theorem}{Theorem}

\begin{document}
\begin{frontmatter}

\title{Synthesis of sup-interpretations: a survey}
\author{Romain P\'echoux}
\address{Université de Lorraine and INRIA team Carte, \\
LORIA, Campus Scientifique - BP 239 - 54506 Vandoeuvre-lès-Nancy Cedex}
\ead{Romain.Pechoux@loria.fr}

\begin{abstract} 
In this paper, we survey the complexity of distinct methods that allow the programmer to synthesize a sup-interpretation, a function providing an upper-bound on the size of the output values computed by a program. It consists in a static space analysis tool without consideration of the time consumption. Although clearly related, sup-interpretation is independent from termination since it only provides an upper bound on the terminating computations. First, we study some undecidable properties of sup-interpretations from a theoretical point of view. Next, we fix term rewriting systems as our computational model and we show that a sup-interpretation can be obtained through the use of a well-known termination technique, the polynomial interpretations. The drawback is that such a method only applies to total functions (strongly normalizing programs). To overcome this problem we also study sup-interpretations through the notion of quasi-interpretation. Quasi-interpretations also suffer from a drawback that lies in the subterm property. This property drastically restricts the shape of the considered functions. Again we overcome this problem by introducing a new notion of interpretations mainly based on the dependency pairs method. We study the decidability and complexity of the sup-interpretation synthesis problem for all these three tools over sets of polynomials. Finally, we take benefit of some previous works on termination and runtime complexity to infer sup-interpretations.
\end{abstract}


\end{frontmatter}

\section{Introduction}
\subsection{Motivations}
The notion of sup-interpretation was introduced in~\cite{MP09} in order to study program extensional complexity. This tool is devoted to statically analyze the complexity of programs guaranteeing that a secured system resists to buffer-overflows and thus allowing the programmer to verify complexity properties of programs used in safety-critical systems. Sup-interpretations focus on analyzing the complexity of programs or, more specifically, term rewrite systems by considering  upper bounds on the size of values computed by a program, by static analysis.\\
Basically, a sup-interpretation of a program is a function that provides an upper-bound on the size of the computed output with respect to the input size. In other words, given a program $\p$, the sup-interpretation of $\p$ is a function that, given some input data $x$ such that $\p$ converges on input $x$, provides an upper-bound on the output size in the size of $x$.\\
One of the main issues concerning static analysis tools is related to their decidability and/or complexity. In other words, one tries to find if the static analysis is decidable and, if so, one tries to study its complexity. As highlighted by Rice's theorem, most of interesting (or non-trivial) analyses are undecidable and, in most of the cases, this issue is transformed into finding the complexity of a smaller instance of the initial problem. In the particular case of sup-interpretations, the analysis consists in finding the sup-interpretation of a given program, that is in synthesizing a function providing upper bounds on the program computations. We call this analysis the \emph{sup-interpretation synthesis problem}. This paper will be dedicated to survey the results concerning the sup-interpre\-tation (SI) synthesis problem. \\

\subsection{Contribution.}
The reader is assumed to be familiar with basic knowledge about term rewrite systems, see chapter 2 of~\cite{Terese} or~\cite{BaaderNipkow98}, and computability and complexity, see~\cite{Rogers87,JonesCC}.\\
We start to show that the general problems of the sup-interpre\-tation synthesis are undecidable when we consider functions and Gödel numberings. Moreover we show that the \emph{sup-interpretation verification problem}, which consists in checking that a function given as input is a sup-interpretation, is $\Pi^0_1$-complete in the arithmetical hierarchy and that the sup-interpretation synthesis problem is in $\Sigma^0_3$.\\
Next we specify our language by introducing Term Rewriting Systems (TRS) and we define the corresponding notion of sup-interpretation. Starting from here, we will study well-known termination and complexity tools like polynomial interpretations (PI) and quasi-interpretations (QI) and show that they allow the programmer to obtain a sup-interpretation under some slight restrictions. \\
We demonstrate that (polynomial) interpretations for termination are special kind of sup-interpretations. However they were designed to study strong normalization and, consequently, they do not provide enough power to study programs computing partial recursive functions. \\
To overcome this problem, we study the notion of quasi-interpretation. We also show that quasi-interpretations define sup-interpretations. \\
Finally, we study a new notion called DP-interpretation (DPI) based on the dependency pairs framework by Arts and Giesl~\cite{AG00} that also defines a sup-interpretation. We show that this new notion strictly generalizes the notion of quasi-interpretation since it does not require any subterm property, a property stating that considered interpretations have to be greater than each of their arguments. In other words, every program admitting a quasi-interpretation admits a DP-interpretation but the converse does not hold. \\
We study the sup-interpretation synthesis problem with respect to each of these tools on particular sets of polynomials ranging over a structure $\bK \in  \{\bN, \bQ ,\bR \}$. The considered sets of polynomials are:
\begin{itemize}
\item the set $\bK[\overline{X}]$ of usual multivariate polynomials whose coefficients are in $\bK$ and with $n$ variables $\overline{X}=\many{X}{n}$ ranging over the field of real numbers, 
\item the set of $\kdMaxPoly\left\{\bK\right\}$ polynomials, which consist in functions obtained using constants over $\bK$ and arbitrary compositions of the operators $+$,$\times$ and $\emph{\max}$ of degree bounded by $\mathsf{d}$ and max arity bounded by $\mathsf{k}$,
\item  and the set of ${\kdMaxPlus}\left\{\bK\right\}$ functions, which consist in functions obtained using constants over $\bK$ bounded by $\mathsf{d}$ and arbitrary compositions of the operators $+$ and $\emph{\max}$, with a max arity bounded by $\mathsf{k}$.
\end{itemize}
The obtained results can be summarized by the following Figure:

\begin{figure}[!h]
\begin{gather*}
\begin{tabular}{|c|c|c|c|}
\hline
 Function space $\backslash$ tool & PI  & QI  & DPI  \\[2pt] \hline
\hline 
& & & \\
$\bK[\overline{X}], \bK \in \{\bN,\bQ\}$  &\text{Undecidable}& \text{Undecidable} &\text{Undecidable}\\[2pt]
\hline
& & & \\
$\bR[\overline{X}]$  & \text{Exptime} & \text{Exptime} & \text{Exptime}\\[2pt]
\hline
& & & \\
$\MaxPoly\{ \bK\}, \bK \in \{\bN,\bQ\}$  & $\maltese$  &\text{Undecidable}  & \text{Undecidable}\\[2pt]
\hline 
& & & \\
$\kdMaxPoly\{ \bR\}$  & $\maltese$ &  \text{Exptime} & \text{Exptime}\\[2pt]
\hline 
& & & \\
$\kdMaxPlus\{ \bK\}, \bK \in \{\bN,\bQ_{\mathsf{d}}\}$  & $\maltese$ & $\NP$-complete  &  $\NP$-complete\\[2pt]
\hline 
& & & \\
$\kdMaxPlus\{ \bR\}$  &  $\maltese$ & $\NP$-hard & $\NP$-hard \\
\hline 
\end{tabular}
\end{gather*}
\label{mondo}
\caption{Decidability and complexity of the sup-interpretation synthesis problem}
\end{figure}

where $\bQ_{\mathsf{d}}$ consists in rationals of bounded representation.\\
The first line is direct consequences of Hilbert's tenth problem undecidability whereas the second line is a consequence of Tarski's quantifier elimination Theorem over real numbers. One important point to mention here is that the synthesis problem is exponential and not doubly exponential because the synthesis problem is more restricted than general quantifier elimination.\\
In the first column, the symbol $\maltese$ means that it does not make sense to study the synthesis problem with respect to the considered set of functions. Indeed the synthesis of polynomial interpretation has no meaning for any structure including a max operator since $\emph{\max}$ is not a strictly monotonic function whereas polynomial interpretations deal with functions enjoying such a property.\\
The results for $\MaxPoly$ function space are identical to the results on pure polynomials since the max operator can be eliminated for both QI and DPI.\\
Finally, in the last two lines of Figure~\ref{mondo}, we show that the synthesis problem is $\NP$-hard for $\MaxPlus$, independently of the structure. As a corollary, on bounded search spaces like $\bN$ or $\bQ_{\mathsf{d}}$, the problem is $\NP$-complete. The meaning of such a notion is unclear over an unbounded and uncountable space like $\bR$. Note that these results are a Corrigendum to results already presented in an unpublished workshop~\cite{BMMP05} that were wrongly stating a $\NP$-completeness result over $\bR$.

Finally, we take benefit of termination results on the runtime complexity of TRS to infer sup-interpretations in a last section. In analogy with complexity theory, we show that time bounded computations imply size (or space) bounded computations. However the space bound may be exponential in the time, if the derivation length is the considered measure of time. Indeed, a derivation of length $n$ may correspond to exponential space by just using variable duplication. We discuss the complexity of the synthesis problem for all of these termination techniques.

\subsection{Outline}
In Section~\ref{S2} we consider general undecidable problems of the sup-interpre\-tation synthesis when considering functions. In Section~\ref{S3}, we introduce Term Rewriting Systems and the corresponding notion of sup-interpretation that slightly differs from the sup-interpretation on functions. In Section~\ref{S4}, we introduce polynomial interpretations as sup-interpretations and study the decidability and complexity of their sup-interpre\-tation synthesis problem. Sections~\ref{S5} and~\ref{S6} apply the same analysis to the notions of quasi-interpretation and DP-interpretation. Section~\ref{S7} discusses the relation between time and space, where time is considered to be the derivation length and space is considered to be the size of a term. This section shows how to synthesize a sup-interpretation through the use of termination techniques. Finally, Section~\ref{S8} discusses the main open issues.

\subsection{Related works}
Sup-interpretations are inspired by two former notions on Term Rewriting Systems, the polynomial interpretations, introduced in~\cite{Lankford,MN70} to analyze program termination and runtime complexity~\cite{CL92,HL88,MS08}, and the quasi-interpretations, introduced in~\cite{MM00} and used to characterize complexity classes such as $\FPtime$, $\FPspace$ or $\Logspace$ (See~\cite{BD10,BMM01,BMM05}). \\
The general framework of sup-interpretation was introduced in~\cite{MP09} without considering the synthesis problem. \cite{MP08} was the first paper to combine interpretation methods together with the dependency pairs method in order to characterize polynomial time and space complexity classes in a more intensional way, that is by capturing more natural algorithms corresponding to a given polynomial time or space function. However the results were presented independently of the notion of sup-interpretation and the present paper gives a deeper understanding on the combination of both methods in order to obtain a sup-interpretation.\\
One important point to stress here is that sup-interpretations are an extensional tool contrarily to quasi-interpretations and polynomial interpretations that are intensional tools. It means that sup-interpretations deal with functions as mathematical object in the sense of complexity theory, that is functions computed by some programs, whereas (general) interpretations are intensional tools and deal with program properties. As a consequence, they also allow the programmer to study finer and more technical program behaviors. For example, a quasi-interpretation also provides upper-bounds on the size of a program intermediate computations whereas this property has no meaning for a sup-interpretation. However sup-interpretations can be combined in criteria in order to get intensional properties such as upper bounds on the size of intermediate values. The aim of this paper is neither to cover the way to get such intensional properties nor to show how they can help in characterizing complexity classes. Consequently, we encourage the interested reader to study~\cite{MP09}.\\
The paper~\cite{Amadio03} has already deeply studied the synthesis problem for quasi-interpretations using max-polynomials with additive coefficient in $\bN$ or $\{0,1\}$ and variables in $\bQ$. The present work takes advantage of these results to present them from a sup-interpretation point of view. Moreover they are extended, firstly, by considering rational and real multiplicative coefficients and, secondly, by extending the $\NP$-hardness proof of~\cite{Amadio03} over $\bN$ to $\NP$-completeness results over natural numbers and rational numbers of bounded representation (and not only $\{0,1\}$).
One last and important point is that the aim of the current paper is not to provide an automated way to synthesize a sup-interpretation but to find the complexity of the synthesis problem depending on the tool used (interpretation, quasi-interpretation, DP-interpretation, termination tools...), on the set of considered functions (polynomials, polynomials with max,...) and on the considered domain (positive real or rational numbers, natural numbers,...). The reader interested by automation should refer to the recent papers~\cite{F07,F08,B11} that allow to build interpretations (and consequently, sup-interpretations as demonstrated in Section~\ref{S4}) for showing program termination and to the tools that synthesize quasi-interpretations~\cite{AMS08,BMP07}. 

\section{Undecidability results}\label{S2}
In this section, we show undecidability results for the synthesis of sup-interpretations. All these results are machine independent and rely on simple Cantor's diagonalizations using Gödel numbering and $s^m_n$ theorem:
\begin{definition}
Suppose that we have a fixed procedure that lists all the sequences of instructions. It associates the set of instructions $P_x$, the $(x+1)st$ set of instructions in the list, to each integer $x$. $x$ is called the Gödel number of $P_x$ and it corresponds to the partial recursive function $\varphi_x$ determined by $P_x$.
\end{definition}
\begin{theorem}[Kleene $s^m_n$~\cite{Kleene88}]
$\forall m,n \geq 1$ there is a recursive function $s_n^m$ of arity $m+1$ such that $\forall x,y_1,\ldots,y_m$:
$$\lambda z_1. \ldots \lambda z_n. \varphi_x(y_1,\ldots,y_m,z_1,\ldots,z_n)=\varphi_{s_n^m(x,y_1,\ldots,y_m)}$$
\end{theorem}
In what follows, let $PRF$ be the set of partial recursive functions and $RF$ be the set of total recursive functions of domain and codomain $\bN$. Given a function $f \in PRF$ and some number $x$, we write $f(x) \downarrow$ (respectively $\downarrow_t$) if $f$ yields an output on input $x$ (resp. in time $t$), and we write $f(x) \uparrow$ otherwise (resp. $\uparrow_t$ otherwise). Consequently $f(x) \downarrow$ is equivalent to $\exists t,\ f(x) \downarrow_t$.
$\mu$ is the classical minimization operator. Given a property $P(x)$, $\mu x.P(x)$ is the smallest $x$ satisfying $P$.
\begin{definition}
Given a function $f \in PRF$, a sup-interpre\-tation of $f$ is a function $F \in RF$ that bounds $f$ on its definition domain, i.e.~$\forall x \in \mathbb{N},\ f(x) \downarrow \implies F(x) \geq f(x)$.
\end{definition}
First we can show as a direct consequence of Rice's Theorem that there exist partial recursive functions that do not have any recursive sup-interpretation:
\begin{theorem}
$$\neg (\forall f \in PRF,\ \exists F \in RF, \forall x \in \bN, f(x)\downarrow \implies F(x) \geq f(x))$$
\end{theorem}
\begin{proof}
Suppose that the implication $\forall f \in PRF,\ \exists F \in RF,\ \forall x \in \bN, f(x)\downarrow \implies F(x) \geq f(x)$ holds.
Define the function $f$ by $f(n)=\varphi_n(n)+1$. By definition, $f$ is clearly in $PRF$. Consequently, $\exists F \in RF$ such that $\forall x \in \bN, f(x)\downarrow \implies F(x) \geq f(x)$. Let $i$ be the Gödel number of such a function $F$. We obtain that $\forall x \in \bN, f(x)\downarrow \implies \forall x,\ \varphi_i (x) \geq f(x)$. As a consequence, $\varphi_i(i)  \geq f(i)=\varphi_i(i)+1$. It contradicts the hypothesis that $F \in RF$.
\end{proof}
This diagonalization result no longer holds if we allow $F$ to be in $PRF$ (In this case, we can trivially set $F(x)=f(x)$). \\
Now we try to find a recursive function that given two Gödel numbers $x$ and $y$ would allow us to compare the corresponding partial recursive functions $\varphi_x$ and $\varphi_y$. We also obtain a negative answer to this issue. 
\begin{theorem}\label{verif}
$\not\exists F \in RF,$
$$F(x,y) = \left\lbrace \begin{array}{l}
1 \quad\text{if }\forall z,\ \varphi_{x}(z)\downarrow \Longrightarrow  \varphi_{x}(z) \leq \varphi_y(z) \\
0 \quad\text{otherwise}
 \end{array}
\right.$$
\end{theorem}
\begin{proof}
Suppose that such a recursive function $F$ exists and define $f$ to be the characteristic function of $\{<x,y>| \forall z,\ \varphi_x(z) \downarrow \implies \varphi_x(z) \leq \varphi_y(z) \}$. $f$ is recursive. Define $\phi$ to be a function of two variables corresponding to the following instructions set: given the  input $<x,y>$, apply $P_x$ to $x$ and return $0$ if and when this computation converges. By Church-Turing thesis, it defines a partial recursive function:
$$\phi(x,y) = \left\lbrace \begin{array}{l}
0 \quad \text{if }\varphi_{x}(x)\downarrow \\
\uparrow \quad \text{otherwise}
 \end{array}
\right.$$
Suppose that $i$ is the Gödel number of such a function, applying $s^m_n$ Theorem, we obtain that there is a recursive function $s^1_1$ such that $\forall x,\ \lambda y.\phi(x,y)=\varphi_{s^1_1(i,x)}$.\\
Now suppose that $x_0$ is a Gödel number for the constant function $\lambda x.0$. We have that $\lambda x. f(s^1_1(i,x),x_0)$ is recursive since it is obtained by composition of recursive functions. However by definition:
\begin{align*}
 f(s^1_1(i,x),x_0) &= \left\lbrace \begin{array}{l}
1 \quad\text{ if }\forall z,\ \varphi_{s^1_1(i,x)}(z)\downarrow \Longrightarrow  \varphi_{s^1_1(i,x)}(z) \leq 0 \\
0 \quad\text{ otherwise}
 \end{array}
\right.\\
&= \left\lbrace \begin{array}{l}
1 \quad\text{ if }\varphi_{s^1_1(i,x)}(z)= 0 \\
0 \quad\text{ otherwise}
 \end{array}
\right.\\
&= \left\lbrace \begin{array}{l}
1 \quad\text{ if } \varphi_{x}(x)\downarrow \\
0 \quad\text{ otherwise}
 \end{array}
\right.
\end{align*}
So we have reduced our function to a variant of the halting problem (see Rogers~\cite{Rogers87}) which is known to be undecidable. Consequently, $\lambda x. f(s^1_1(i,x),x_0)$ is not recursive and we obtain a contradiction. 
\end{proof}
Consequently, we obtain that the \emph{sup-interpretation verification problem} defined by $ SI(F)=\{ x \ | \ \forall z\ \varphi_x(z) \downarrow \implies \varphi_x(z) \leq F(z) ) \}$, which consists in checking that a given function $F$ is a sup-interpretation of a function $f$ of index $x$ (i.e. $x \in SI(F)$), is undecidable. As a corollary, we also obtain that the \emph{sup-interpretation synthesis problem}, which consists in finding the smallest function wrt Gödel numbering that bounds another given as input, is also undecidable:
\begin{corollary}\label{min}
$\not\exists G \in RF$ such that:
$$G(x) = \left\lbrace \begin{array}{l}
\mu y. \{\forall z,\ \varphi_{x}(z)\downarrow \Longrightarrow \varphi_{x}(z) \leq \varphi_{y}(z)\}\\
0 \quad\text{ otherwise}
 \end{array}
\right.$$
\end{corollary}
\begin{proof}
Assume that $G$ is recursive and that we have a Gödel numbering starting from Gödel number $1$ (i.e.~not defined in $0$). The reason for which we take such a numbering is just that we do not want to make a confusion between the output $0$ when there is no upper-bound and the index $0$ of the function $\varphi_0$ that might be an upper bound of some other function. Then $\mu y.F(x,y)$, with $F$ defined in Theorem~\ref{verif} has the same characteristic function than $G$. Consequently, we obtain a contradiction and $G$ cannot be recursive.
\end{proof}
Now let just state that the sup-interpretation verification problem which consists in checking that a fixed function $F$ is a sup-interpretation of a function $\varphi_x$ of index $x$, noted $SI(F)$ is $\Pi^0_1$-complete in the arithmetical hierarchy:
\begin{theorem}
The sup-interpretation verification problem $SI(F)$ is $\Pi^0_1$-complete.
\end{theorem}
\begin{proof}
For every input $z$ and every $t$, either $\varphi_x(z)$ terminates within time $t$ and, in this case, we have to compare $\varphi_x$ and $F$ or $\varphi_x(z)$ does not terminate in time $t$. Consequently, we can write
$ SI(F)=\{ x \ | \ \forall z, \ \forall t,\ \varphi_x(z) \uparrow_t \vee (\varphi_x(z) \downarrow_t \wedge \varphi_x(z) \leq F(z) ) \}$. \\
We briefly recall that a problem $B$ is complete for some class $C$ of the arithmetical hierarchy if there is a total computable function $f$ such that:
$$x \in A \text{ iff } f(x) \in B$$
for some problem $A$ known to be $C$-complete. Consider the problem $A=\{x \ | \ \varphi_x(0) \uparrow\}$. This problem is known to be $\Pi^0_1-complete$ since it is co-RE.\\
Now define the function $f$ such that for each $x$ the function $\varphi_{f(x)}$ of index $f(x)$ is defined by:
$$\varphi_{f(x)}=\begin{cases} F(z)+1 \quad \text{ if } \varphi_x(0) \downarrow \\ \uparrow \quad \text{otherwise} \end{cases}$$
We clearly have:
$$ x \in A \text{ iff } f(x) \in SI(F)$$
Moreover the function $f$ is clearly total, by definition, and computable, by applying $s^m_n$ Theorem. Consequently, $SI(F)$ is $\Pi^0_1-complete$.
\end{proof}

Now we show that sup-interpretation synthesis problem, $SI$ defined to be ``the set of functions $f \in PRF$ for which there is a total recursive function $F$, satisfying: for all $z \in \bN$ if $f(z) \downarrow$ then $F(z) \geq f(z)$'' is $\Sigma^0_3$ in the arithmetical hierarchy:
\begin{theorem}
$SI \in\Sigma^0_3$.
\end{theorem}
\begin{proof}
$SI$ can be written equivalently as:
$$SI=\{ x \in \bN \ | \ \exists s, \forall z,\exists t, \varphi_s(z) \downarrow_t \wedge ( \varphi_x(z) \downarrow_t \implies \varphi_s(z) \geq \varphi_x(z)) \} $$
In other words, $SI$ is the the set of indexes $x$ corresponding to functions $\varphi_x$ for which there exists a total function $\varphi_s$ providing an upper bound on terminating computations (Indeed $ \varphi_x(z) \downarrow_t \implies \varphi_s(z) \geq \varphi_x(z)$). The formula $ \varphi_s(z) \downarrow_t \wedge ( \varphi_x(z) \downarrow_t \implies \varphi_s(z) \geq \varphi_x(z)) \in \Pi_0^0$ and, consequently, $SI \in \Sigma^0_3$.
\end{proof}

\section{Sup-interpretations over Term Rewriting Systems}\label{S3}
\subsection{TRS as a computational model}
The previous section only deals with machine independent results and we have hidden for a while the data representation problems arising. Consequently, we have to adapt slightly the notion of sup-interpre\-tation to each computational model under consideration.
Throughout the following Sections, we will consider term rewriting systems.\\ 
A Term Rewriting System (TRS for short) is a formal system for manipulating terms over a signature by means of rules.\\ 
Terms are strings of symbols consisting of a countably infinite set of variables $\Var$ and a first-order signature $\Sigma$, a non-empty set of function symbols or operator symbols of fixed arity. $\Var$ and $\Sigma$ are supposed to be disjoint. As usual, the notation $ \Terms{\Sigma,\Var}$ will be used to denote the set of terms $s,t,\ldots$ of signature $\Sigma$ and having variables in $\Var$. \\
A (one-hole) context $C[\diamond]$ is a term in $\Freetermsd$ with exactly one occurrence of the hole $\diamond$, a symbol of arity $0$. Given a term $t$ and context $C[\diamond]$, let $C[t]$ denote the result of replacing the hole $\diamond$ with the term $t$.\\
A substitution $\sigma$ is a mapping from $\Var$ to $\Freeterms$.\\
A rewrite rule for a signature $\Sigma$ is a pair $l \to r $ of terms $l,r \in \Freeterms$.
A Term Rewrite System is as a pair $\proga$ of a signature $ \Sigma$ and a set of rewrite rules $\Regles$. In what follows, we will suppose that all the variables of a right-hand side $r$ are included in the variables of $l$ as in Chapter 2 of~\cite{Terese}. \\
A constructor Term Rewrite System is a TRS in which the signature $\Sigma$ can be partitioned into the disjoint union of a set of function symbols $\Functions$ and a set of constructors $\Cns$, such that for every rewrite rule $l \to r$ we have $l=\funone(\many{t}{n})$ with $\funone \in \Fct$ and $\many{t}{n} \in \Terms{\Cns,\Var}$. The constructors are introduced to represent inductive data. They basically consist of a strict subset $\Cns \subset \Sigma$ of non-defined functions (a function is defined if it is the root of a left-hand side term in a rule). In what follows, we will only consider constructor TRS and we will use the notation $\prog$ to denote such a particular TRS, $\Cns \uplus \Fct$ being the disjoint union of the sets $\Cns$ and $\Fct$. Terms in $\Terms{\Cns,\Var}$ will be called patterns. \\
In what follows, we will consider orthogonal constructor TRS since we only want to deal with functions. The notion of orthogonality requires that reduction rules of the system are all left-linear, that is each variable occurs only once on the left hand side of each rule, and there is no overlap between patterns. It is a sufficient condition to ensure that the considered TRS is confluent. It implies that we are clearly talking of functions that maps a term to another (and not functions mapping a term to a set of terms in the case of non-confluent systems). Note that this syntactic requirement could have been withdrawn in favor of a semantic restriction that would only consider TRS that compute functions. Our choice restricts the expressivity of considered TRS but makes sense in our theoretical development since it does not restrict the computed functions set. \\
Given two terms $s$ and $t$, we have that $s \to_\Regles t$ if there are a substitution $\sigma$, a context $C[\diamond]$ and a rule $l \to r \in \Regles$ such that $s=C[l\sigma]$ and $t=C[r\sigma]$. Throughout the paper, let $\to_{\Regles}^*$ (resp.~$\to_{\Regles}^+$) be the reflexive and transitive (resp.~transitive) closure of $\to_{\Regles}$. Moreover we write $s \to_\Regles^n t$ if $n$ rewrite steps are performed to rewrite $s$ to $t$. A TRS terminates if there is no infinite reduction through $\to_{\Regles}$.\\
A function symbol $\funone$ of arity $n$ will define a partial function $\sem{\funone}$ from constructor terms\footnote{As usual $\Consterms = \Terms{\Cns,\emptyset}$} (sometimes called values) $\Consterms^n$ to $\Consterms$ by:
$$\forall \many{v}{n} \in \Consterms,\ \sem{\funone}(\many{v}{n})=v \text{ iff } \funone(\many{v}{n}) \to_{\Regles}^* v \ \wedge \ v \in \Consterms$$ 
In this case, we write $\sem{\funone}(\many{v}{n}) \downarrow$ to mean that the computation ends in a normal form (constructor term). If there is no such a $v$ (because of divergence or because evaluation cannot reach a constructor term), then  $\sem{\funone}(\many{v}{n}) \uparrow$. Finally, we define the notion of size of a term $\taille{\termone}$ which is equal to the number of symbols in $\termone$.

\subsection{Sup-interpretation of a TRS}
Since the goal of sup-interpretation is to provide a non-negative upper bound on the size of computed values, we will mainly restrict our analysis to the groups $\bN, \bQ$ and $\bR$, where $\bQ$ and $\bR$ denote positive rational numbers and positive real numbers. In what follows, let $\bK \in \{\bN,\bQ,\bR\}$ and let $\geq$ and $>$ be the natural ordering and strict ordering on such a structure. Finally, let $>_\delta$ be the strict ordering defined by $x>_\delta y$ iff $ x \geq \delta+y$, for some fixed $\delta \in \bK$ such that $\delta >0$.
\begin{definition}\label{ass}
Given a TRS $\prog$, an assignment $\theta$ over $\bK$ is a mapping that maps every symbol $\funtw \in \Fct \uplus \Cns$ of arity $m$ to a total function $\theta(\funtw) : \bK^m \to \bK$ and that maps every variable $\in \Var$ to a variable in $\bK$.\\
 An assignment is \emph{additive} if $\forall c \in \Cns$ of arity $n>0$,  $\theta(c)=\lambda \many {x}{n}.(x_1+\ldots +x_n +k_c)$, for some $k_c \geq 1$, and $\forall c \in \Cns$ of arity $0$, $\theta(c)=0$.  An assignment is k-\emph{additive} if for all $c \in \Cns$, $k_c\leq k$.
\end{definition}
\begin{definition}
An assignment $\theta$ over $\bK$ is (strictly) monotonic if for every symbol $\funone$ of arity $m$, $\theta(\funone)$ is a (strictly) monotonic function in each of its arguments. In other words, $\forall i \in [1,m],\ x\geq y \implies \theta(\funone)(\ldots,x_{i-1},x,x_{i+1},\ldots) \geq \theta(\funone)(\ldots,x_{i-1},y,x_{i+1},\ldots)$ (resp.~$\forall i \in [1,m],\forall \delta >0,\ \exists \epsilon >0,\ \theta(\funone)(\ldots,x+\delta,\ldots) >_\epsilon \theta(\funone)(\ldots,x,\ldots)$).
\end{definition}
Now we are able to adapt the notion of sup-interpretation to this model:
\begin{definition}[Sup-interpretation]\label{si}
Given a TRS $\prog$, a monotonic and additive assignment $\theta$ over $\bK$ is a sup-interpre\-tation over $\bK$ if $\forall \funone \in \Fct$ of arity  $m$ and $\forall \many{v}{m} \in \Consterms$:
$$\funone(\many{v}{m}) \downarrow \implies \\ \theta(\funone(\many{v}{m})) \geq \theta(\sem{\funone}(\many{v}{m}))$$
where the sup-interpretation $\theta$ is extended canonically to general terms by:
$$\theta(\funtw(\eone_1,\ldots,\eone_n))= \theta(\funtw)(\theta(\eone_1),\ldots,\theta(\eone_n)), \quad \funtw \in \Fct \uplus \Cns $$
\end{definition}
We restrict the shape of constructor symbol sup-interpretations by requiring a $k$-additive assignment. This restriction is made to relate easily the interpretation of a constructor term and its size, i.e.  $\exists k \in \bN,\ \forall v \in \Consterms, k \times \taille{v}\geq \theta(v) \geq \taille{v}$ always hold for a TRS wrt a fixed additive sup-interpretation.\\
We compare this new definition wrt the one presented in previous Section: in a given TRS, the sup-interpretation of a function symbol $\funone$ of arity $m$ can be discretized to be viewed as a function  $\theta(\funone): \bN^m \to \bN$ that bounds the size of the output wrt to the input sizes (this is direct for 1-additive sup-interpretations):
\begin{lemma}
Given a TRS $\prog$ having a sup-interpretation $\theta$ then for each function symbol $\funone \in \Fct$ and for all values $\many{v}{m} \in \Consterms$ such that $\funone(\many{v}{m}) \downarrow$, we have:
$$ \theta(\funone)(k\times \taille{v_1},\ldots,k\times\taille{v_m})  \geq \taille{\sem{\funone}(\many{v}{m})}$$
\end{lemma}
\begin{proof}
\begin{align*}
&\theta(\funone)(k\times \taille{v_1},\ldots,k\times\taille{v_m}) &&\text{By monotonicity} \\
&\geq \theta(\funone)(\ttmany{v}{m}) &&\text{and }k\text{-additivity}\\
&= \theta(\funone(\many{v}{m}))&&\text{By extension}\\
&\geq \theta(\sem{\funone}(\many{v}{m}))&&\text{By Definition\ \ref{si}}\\
&\geq \taille{\sem{\funone}(\many{v}{m})}&&\text{By }k\text{-additivity}
\end{align*}
and so the conclusion.
\end{proof}

\section{Polynomial interpretations}\label{S4}
\subsection{Interpretations as sup-interpretations}
Given a TRS, the main issue is now to synthesize a sup-interpreta\-tion, that is to compute an upper-bound on the partial function it computes. The first natural technique to do so comes from the term rewriting termination community, is called (polynomial) interpretation and was introduced in~\cite{MN70, Lankford}.
\begin{definition}[Interpretation]\label{i}
Given a TRS $\prog$, an (additive) interpretation is a \emph{strictly} monotonic (additive) assignment $[-]$ over $\bK$ which satisfies:
\begin{enumerate}
\item $\forall l \to r \in \Regles, \ [l] > [r]$
\item If $\bK \in \{ \bQ, \bR \}$ then:
\begin{enumerate}
\item either $\forall \funtw \in \Fct \uplus \Cns,$ of arity $m>0$,\\ $\forall i \in [1,m],\ [\funtw](\many{X}{m})>X_i$
\item or $\forall l \to r \in \Regles, \ [l] >_\delta [r]$
\end{enumerate}
\end{enumerate}
where the interpretation $[-]$ is extended canonically to terms as usual.
\end{definition}
Condition 1 constitutes the basis of interpretation method as introduced in~\cite{MN70, Lankford}. Condition 2(a) was introduced by Dershowitz~\cite{D79} to compensate for the loss of well-foundedness over the reals. Finally, condition 2(b) is due to Lucas~\cite{Lucas06} and captures more TRS than 2(a).\\
As demonstrated in~\cite{Lankford}, an interpretation defines a reduction ordering (i.e.~a strict, stable, monotonic and well-founded ordering) 
\begin{theorem}\label{terminaison}
If a TRS $\prog$ admits an interpretation then it terminates.
\end{theorem}
Moreover, an additive interpretation defines a sup-interpre\-tation:
\begin{theorem}\label{iissi}
Given a TRS $\prog$ having an additive interpretation $[-]$ then $[-]$ is a sup-interpretation.
\end{theorem}
\begin{proof}
First note that the assignment is additive by assumption.\\
Second, we show that for each values $\many{v}{n} \in \Consterms$ and function symbol $\funone \in \Fct$ such that $\funone(\many{v}{n}) \downarrow$ we have $[\funone(\many{v}{n})] \geq [\sem{\funone}(\many{v}{n})]$.
Consider a function symbol $\funone$ and values $\many{v}{n}$, by Theorem~\ref{terminaison}, we have $\funone(\many{v}{n}) \downarrow$. Since interpretations define a reduction ordering, we have that each reduction corresponds to a (strictly) decreasing sequence:
\begin{align*}
&\funone(\many{v}{n}) \to_{\Regles} u_1 \to_{\Regles} \ldots \to_{\Regles} u_k \to_{\Regles} \sem{\funone}(\many{v}{n})\\
&[\funone(\many{v}{n})] > [u_1] > \ldots > [u_k] > [\sem{\funone}(\many{v}{n})]
\end{align*}
and, a fortiori,  $[\funone(\many{v}{n})] \geq [\sem{\funone}(\many{v}{n})]$.
\end{proof}
Consequently, finding the interpretation of a given program provides a sup-interpretation of this program under additivity constraints as illustrated by the following example:
\begin{example} Consider the following simple TRS:
\begin{align*}
&\double(0) \to 0 &&\sexp(0) \to 1\\
&\double(x+1) \to \double(x)+2 &&\sexp(x+1) \to \double(\sexp(x))
\end{align*}
where $x+2$ and $1$ are notations for $(x+1)+1$ and $0+1$. It admits the following additive interpretation $[0]=0$, $[+1](X)=X+1$, $[\double](X)=3\times X+1$, $[\exp](X)=3^{2\times X+1}$. Indeed, it is a strictly monotonic additive assignment and for the last rule, we have:
\begin{align*}
[\sexp](x+1) &=3^{2\times [(x+1)]+1} =3^{2(X+1)+1}=3^{2X+3}\\
&>  3 \times 3^{2X+1}+1 =[\double](3^{2X+1})=[\double]([\sexp(x)])
\end{align*}
We let the reader check that the strict inequalities hold for the other rules.
\end{example}

\subsection{Restriction to polynomials}
It is natural to restrict the space of considered functions (the sup-interpreta\-tion codomain) to polynomials for two reasons. First, as we have seen in the first Section, considering the whole space of functions is too general in terms of decidability. Second, polynomials are admitted to be a relevant set of functions in term of time and space complexity. Consequently, we restrict the function space in order to get effective procedures.\\
In what follows, let $\bK[X_1,\ldots,X_m]$ be the set of $m$-ary polynomials whose coefficients are in $\bK$.
\begin{definition}[Polynomial interpretation]\label{PID}
Given a TRS $\prog$, a polynomial interpretation over $\bK$ is an interpretation $[-]$ over $\bK$ that maps every symbol $\funtw \in \Fct \uplus \Cns$ of arity $m$ to a function $[\funtw] \in \bK[X_1,\ldots,X_m]$.
\end{definition}
The synthesis problem for polynomial interpretation has been deeply studied in~\cite{CMTU05,Lucas07} where algorithms solving the constraints are described. More recently, encoding-based algorithms via $\mathsf{SAT}$ or $\mathsf{SMT}$ solving have become the state of the art for the synthesis problem~\cite{F07,B11}. One important question is what is the best structure ($\bN, \bQ$ or $\bR$) to consider in order to get a polynomial interpretation. This question has no answer as surveyed by the following results:
\begin{theorem}[Lucas~\cite{Lucas06}]\label{Luc}
There are TRS that can be proved terminating using a polynomial interpretation over $\mathbb{R}$, whereas they cannot be proved terminating using a polynomial interpretation over $\mathbb{Q}$.
\end{theorem}
\begin{theorem}[Lucas~\cite{Lucas06}]
 There are TRS which can be proved terminating using a polynomial interpretation over $\mathbb{Q}$, whereas they cannot be proved terminating using a polynomial interpretation over $\bN$.
\end{theorem}
\begin{theorem}[Middeldorp-Neurauter.~\cite{NM10}]
 There are TRS which can be proved terminating using a polynomial interpretation over $\bN$, whereas they cannot be proved terminating using a polynomial interpretation over $\mathbb{Q}$ or $\mathbb{R}$.
\end{theorem}

\subsection{Decidability results over polynomials}
However we can compare the structures through decidability or undecidability results for the sup-interpretation synthesis problem as illustrated below.
\begin{definition}[PI synthesis problem]
Given a TRS $\prog$, is there an assignment $[-]$ such that $[-]$ is a polynomial interpretation of  $\prog$?
\end{definition}
\begin{theorem}
 The PI synthesis problem is undecidable over $\bN[\overline{X}]$ and $\bQ[\overline{X}]$.
\end{theorem}
\begin{proof}
This is a direct consequence of Hilbert's tenth Problem undecidability since every inequality of Definition~\ref{PID} of the shape $\forall [\varone_1], \ldots ,[\varone_n], [l] > [r]$, $\many{\varone}{n}$ being the free variables of $l$, can be turned into the satisfaction of the formula $ \neg \exists  [\varone_1], \ldots ,[\varone_n],\  [l] - [r]=0$. The interested reader should refer to~\cite{Matiyasevich93}. Note that we have not checked that each arbitrary polynomial can be encoded. This technical check which is needed to show a reduction from Hilbert's tenth problem to the PI synthesis will be performed in the next section for the notion of quasi-interpretation. 
\end{proof}
This result was historically mentioned to be undecidable by Lankford~\cite{Lankford}.\\

Now we show that the polynomial interpretation synthesis problem is decidable over $\bR$ as a corollary of Tarski's Theorem~\cite{Tarski51}. Historically, Tarski's procedure was non-elementary. It has been improved by Collins~\cite{Collins75} in a procedure of complexity doubly exponential in the number of variables. We will use the most precise upper bound on such a procedure known by the author and described in~\cite{HRS90}, where the procedure is shown to be doubly exponential in the number of quantifiers blocks alternations and exponential in the number of variables, in order to exhibit a precise upper bound on the complexity of the PI synthesis problem: we will obtain an \emph{exponential} procedure because the polynomial quasi-interpretation synthesis problem is more restricted than the general quantifiers elimination over $\bR$ described by Tarski. 
%
\begin{theorem}[Roy et Al.~\cite{HRS90}]\label{roy} Given an integral domain $\mathsf{k}$ (i.e.~a commutative ring with no zero divisor) included in a real closed field $\textbf{R}$,  a formula $\phi$ of size $L$ in the ordered fields language under prenex normal form with parameters in $\mathbf{K}$, containing $m$ blocks of quantifiers and $s$ polynomials of $n$ variables and with coefficients in $\mathsf{k}$ whose sum of degrees is less or equal to $D$, there is an algorithm of complexity $O(L)D^{n^{O(m)}}$ which computes an equivalent quantifier-free formula.
\end{theorem}

\begin{theorem}\label{thm:synthese}
The PI synthesis problem is decidable in exponential time (in the size of the program) over $\bR[\overline{X}]$.
\end{theorem}

\begin{proof}
We start by encoding the strict monotonicity property:
Given a TRS $\prog$, $\funone \in \Fct$ of arity $n$ and an assignment $[-] \in \bR[\overline{X}]$ such that $[\funone]$ is defined,  the strict monotonicity property can be encoded by the following first order formula:
\begin{align*}
SM[\funone] &= \forall X_1, \ldots, X_n,\forall Y_1,\ldots,Y_n,\\  
&\underset{{l \in [1,n]}}{\bigwedge}X_l > Y_l \implies  [\funone](\many{X}{n}) > [\funone](\many{X}{n})
\end{align*}
In other words, $SM[\funone]$ if and only if $[\funone]$ is strictly monotonic.\\
Now we encode the inequalities for each rule of a given program $\prog$:
Given a TRS $\prog$, of assignment $[-]$, let $\overline{a}$ be a enumeration of the multiplicative coefficients involved in the polynomials $[\funone]$, $\forall \funone \in \Fct \uplus \Cns$, and define $PI[\prog] = \exists \overline{a} \in \bR, (\bigwedge_{\funone \in \Functions \uplus \Cns} SM[\funone]) \wedge  (\bigwedge_{l \to r \in {\Regles}}  [l] > [r])$.\\
$PI[\prog]$ is true if and only if there is an assignment $[-]$ that is a  polynomial interpretation of $\prog$.\\
Performing a careful $\alpha$-conversion of all the variables occurring in the distinct inequalities of the formula $PI[\prog]$, we can extrude all the quantifiers (existential and universal) to obtain a new formula under prenex normal form with only one alternation between a block of existential quantifiers (encoding the polynomials multiplicative coefficients) and one block of universal quantifiers (encoding program variables). \\
Now we apply Theorem~\ref{roy} by setting $\mathbf{K}=\mathbb{R}$ and $\phi=PI[\prog]$ and we obtain an algorithm of complexity $O(\taille{PI[\prog]})D^{n^{O(m)}}$ which computes an equivalent quantifier-free formula. Note that:
\begin{itemize}
\item the size of the formula $PI[\prog]$ is bounded polynomially by the size of the program and exponentially by the maximal degree of the polynomial, which is also bounded by $D$. Indeed the number of multiplicative coefficients within a polynomial of bounded degree $D$ is exponential in $D$.
\item the number $n$ of variables is bounded polynomially by the size of the program and exponentially by the degree $D$
\item the number $m$ of blocks is bounded by $2$
\end{itemize}
Consequently, the algorithm has a complexity exponential in the size of the program.
\end{proof}

\subsection{Drawbacks of (polynomial) interpretations}
The previous Subsection has provided a positive result, that is a mechanical way to synthesize the sup-interpretation of a given program. On the other hand, Theorem~\ref{terminaison} can be interpreted as a negative result. Indeed, in terms of TRS, termination means that either the evaluation stops on a constructor term $v \in \Consterms$ or that the evaluation stops on a (undefined) term still containing non-evaluated function symbols in $\Fct$. In particular, it means that this analysis rejects all the partial functions that diverge on some input domain but still remain bounded on its complement, as illustrated by the following example:
\begin{example}\label{qiex}
\begin{align*}
&\funone(x+2) \to \funone(x)+2 &&\funone(0) \to \funone(0) &&&\funone(1) \to 1
\end{align*}
The function $\funone$ computes the identity function on odd numbers whereas it infinitely diverges on even numbers. Consequently, it does not admit any polynomial interpretation whereas we would expect $\theta(\funone)(X)=X$ to be a suitable sup-interpretation.
\end{example}

 \section{Quasi-interpretations}\label{synthese de quasi-interprtations}\label{S5}
\subsection{Quasi-interpretations as sup-interpretations}
We introduce the notion of quasi-interpretation~\cite{BMM07} that, in contrast with (polynomial) interpretations, allows us to study partial functions.
\begin{definition}[Quasi-interpretation]\label{qidef}
Given a TRS $\prog$, a (additive) quasi-interpretation (QI for short) is a monotonic (additive) assignment $\interp{-}$ over $\bK$ satisfying:
\begin{enumerate}
\item $\forall l \to r \in \Regles, \ \interp{l} \geq \interp{r}$
\item $\forall \funtw \in\Fct \uplus \Cns,$ of arity $m$, $\forall i \in [1,m],\ \interp{\funtw}(\many{X}{m}) \geq X_i$
\end{enumerate}
where the quasi-interpretation $\interp{-}$ is extended canonically to terms as usual.
\end{definition}
Condition 2 is called the \emph{subterm property}. 
Quasi-interpreta\-tions do not tell anything about program termination since the strict ordering of Definition~\ref{i} has been replaced by its reflexive closure. Well-foundedness is lost and this is the main reason why such a tool can be adapted to partial functions. With this notion, we obtain a result similar to Theorem~\ref{iissi}.

\begin{theorem}\label{qiissi}
Given a program $\prog$ having an additive quasi-interpretation $\interp{-}$ then $\interp{-}$ is a sup-interpretation.
\end{theorem}
\begin{proof}
The proof is essentially the same as the one in Theorem~\ref{iissi}. Strict inequalities are replaced by non-strict inequalities.
\end{proof}

\begin{example}
The program of Example~\ref{qiex} admits the following additive quasi-interpretation:
$\interp{0}=0,\ \interp{+1}(X)=X+1$ and $\interp{\funone}(X)=X$. Indeed, for the first rule, we check:
\begin{align*}
&\interp{\funone(x+2)}=\interp{\funone}(\interp{(x+1)+1)})=X+2\\ 
&\geq  \interp{\funone(x)}+2=\interp{(\funone(x)+1)+1}
\end{align*}
For the second, rule we clearly have $\interp{\funone(0)} \geq \interp{\funone(0)}$ and, for the last rule, we have $\interp{\funone(1)}=\interp{\funone}(\interp{1}) \geq  \interp{1}$.
\end{example}
\subsection{Quasi-interpretation synthesis problem}
The quasi-interpretation synthesis problem was introduced by Amadio in~\cite{Amadio03} and is prominent in the perspective of practical uses of quasi-interpretation since an algorithm synthesizing a quasi-interpretation of a given program would allow the programmer to automatically perform a static analysis of program resources use on terminating computations. 
 It can be defined as follows:
\begin{definition}[QI synthesis problem]
Given a TRS $\prog$, is there an assignment $\interp{-}$ such that $\interp{-}$ is a quasi-interpretation of  $\prog$?
\end{definition}
This problem is undecidable in the general case where we consider total functions as a consequence of Rice's Theorem and as illustrated by Corollary~\ref{min}. Indeed there is no function (and consequently no program) that for a program index given as input provides the smallest index of a sup-interpretation. Consequently, we have to restrict again the set of considered functions. The immediate candidate is the set of polynomials presented in the previous Section. However we choose to add an extra $\emph{\max}$ function. There are many reasons to do so: firstly, $\emph{\max}$ is the smallest function satisfying the subterm condition. Thus it provides the tightest upper bound that we could expect on a function symbol computation. Secondly, it remains stable for the set of polynomials since the max is always bounded by the sum. Lastly, it was not considered in polynomial interpretations for the only reason that it is not strictly monotonic in each of its arguments (i.e. $x > x' \Rightarrow \emph{\max}(x,y) > \emph{\max}(x',y)$ does not hold in the case where $y>x$ with $x,x',y \in \bK$). We define the set of $\MaxPoly$ functions as follows: 
\begin{definition}
Let $\MaxPoly\left\{\bK\right\}$ be the set of functions obtained using constants and variables ranging over $\bK$ and arbitrary compositions of the operators $+$,$\times$ and $\max$.
\end{definition}
We exhibit a normalization result on such a set of functions showing that $\emph{\max}$ operator can be restricted to the upper most level:
\begin{proposition}[Normalisation]\label{prop:norma}
  Each function $Q \in \MaxPoly\{\bK\}, Q \neq 0$, can be written into the following normal form:
\begin{align*}
  Q(X_1,\ldots, X_n) & = \max (P_1(X_1, \ldots, X_n), \ldots, P_k(X_1,
  \ldots, X_n))
\end{align*}
for some $k \geq 1$ and where $P_i \neq 0$ are polynomials.
\end{proposition}
\begin{proof}
By induction on the structure of $Q$:
\begin{itemize}
\item The base case is when $Q$ is a monomial then $Q=\emph{\max}(Q)$. 
\item If $Q=Q_1+Q_2$ then by induction hypothesis $Q_i=\emph{\max}(\many{P^i}{n_i})$, for $i \in \{1,2\}$, with $P^i_j$ polynomials. Consequently, $Q=\emph{\max}(\many{P^1}{n_1})+\emph{\max}(\many{P^2}{n_2})=\emph{\max}_{j \leq n_1, k \leq n_2}(P^1_j+P^2_k)$ since the $\emph{\max}$ operator can be extruded using rules of the shape $\emph{\max}(Q,R)+P = \emph{\max}(Q+P,R+P)$ and $\emph{\max}(\emph{\max}(P,Q),\emph{\max}(R,S))=\emph{\max}(P,Q,R,S)$.
\item In the same way, if $Q=Q_1 \times Q_2$ then $Q=\emph{\max}_{j \leq n_1, k \leq n_2}(P^1_j \times P^2_k)$ 
\end{itemize}
and so the conclusion.
\end{proof}
Moreover, we show that the satisfaction of an inequality in $\MaxPoly\left\{\bK\right\}$ can be transformed into an equivalent problem over polynomials, that is an inequality over $\MaxPoly\left\{\bK\right\}$ can be turned into a conjunction of disjunctions of inequalities over polynomials:
\begin{proposition}\label{bip}
Given an inequality $Q \geq Q',$ with $Q,Q' \in \MaxPoly\left\{\bK\right\}$ there are two integers $n$ and $m$ and polynomials over $\bK$, $P_i,R_j$ for $i \leq n$, $j \leq m$, such that:
$$Q \geq Q' \text{ iff } \bigwedge_{j \in[1,m]} \bigvee_{i \in[1,n]} P_i \geq R_j $$
\end{proposition}
\begin{proof}
By the previous Proposition, $Q$ and $Q'$ can be written as $\emph{\max}(P_1,\ldots,P_n)$ and $\emph{\max}(R_1,\ldots,R_m)$, for some $n$ and $m$. Consequently:
\begin{align*}
&\emph{\max}(P_1,\ldots,P_n) \geq \emph{\max}(R_1,\ldots,R_m) \\
&\Leftrightarrow \bigwedge_{j \in [1,m]} \emph{\max}(P_1,\ldots,P_n) \geq  R_j \\
 &\Leftrightarrow \bigwedge_{j \in [1,m]}\bigvee_{i \in [1,n]} P_i \geq R_j
\end{align*}
and so the result holds.
\end{proof}

\subsection{Undecidable synthesis over Max-Poly$\{\bN\}$}
As expected, the QI synthesis problem remains undecidable over $\bN$ and $\bQ$:
\begin{theorem}\label{QIUN}
 The QI synthesis problem is undecidable over $\MaxPoly\left\{\bN\right\}$ and $\MaxPoly\left\{\bQ\right\}$.
\end{theorem}
\begin{proof}We demonstrate, using Proposition~\ref{bip}, that the synthesis problem over $\MaxPoly\left\{\bN\right\}$ and $\MaxPoly\left\{\bQ\right\}$ can be turned in the satisfaction of (disjunctions and conjunctions of) inequalities of the shape\footnote{This will be shown explicitly in the next Subsection.}: $$\exists \many{a}{n}\forall \many{x}{m},\  P(\many{a}{n},\many{x}{m}) \geq 0$$ where the $a_i$ represent the multiplicative coefficients of the function symbols quasi-interpreta\-tions and where the $x_j$ represent the program variables quasi-inter\-pretations. Fixing the $a_i$, this problem consists in checking that: $$\forall \many{x}{m},\  P'(\many{x}{m}) \geq 0$$ with $P'(\many{x}{m}) =P(\many{a}{n},\many{x}{m}) $. Now we consider Hilbert's tenth problem that was shown to be undecidable over $\bQ$ (and $\bN$) by  Matijasevich~\cite{M93}. Given a polynomial $P$ of arity $n$, there is no procedure that decides: $$\exists \many{x}{n}, \ P(\many{x}{n})=0$$
Over $\bN$, we have:
\begin{align*}
&\exists \many{x}{n}, \ P(\many{x}{n})=0 \\
&\Longleftrightarrow \neg (\forall \many{x}{n}, \ P(\many{x}{n})^2 > 0 )\\
&\Longleftrightarrow \neg (\forall \many{x}{n}, \ P(\many{x}{n})^2-1 \geq 0 )
\end{align*}
Given a polynomial $P$, having a computable procedure that checks whether $\forall \many{x}{n}, \ P(\many{x}{n})^2-1 \geq 0 $ holds would provide a positive answer to Hilbert's problem (and conversely). As a consequence, we know that there is no such a procedure. Finally, we check (a technical but not difficult fact) that for any polynomial $P$ of arity $n$ we can enforce the interpretation of a $n$-ary symbol $\funone$ to satisfy $\interp{\funone}(\many{x}{n})=P(\many{x}{n})^2$ and $\forall \many{x}{n}, \ P(\many{x}{n})^2 \geq 1 $ adding arbitrary rules to a program (provided that $\interp{\funone} \in \MaxPoly\left\{\bN\right\}$). For simplicity, suppose that we have additive constructors $\conone_n$ of arity $n\in \bN$ and such that $\interp{\conone_0}=0$ and $\interp{\conone_n}(\many{X}{n})=\sum_{i=1}^nX_i+1$ for $n \geq 1$, we can encode every natural number $n$ by $n$ compositions of the shape $\underline{n}=\conone_1(\ldots \conone_1(\conone_0)\ldots)$ and we can encode the identity polynomial by adding the following rule:
$$\id(\varone) \to \id(\id(\varone))$$
One can check that the corresponding inequality constraints its quasi-interpreta\-tion to be equal to $\interp{\id}(X)=X$ over $\bN$.
Moreover we can add arbitrary rules of the shape:
\begin{align*}
\id(\conone_n(\varone,\ldots,\varone)) &\to \funone_n (\varone)\\
\id(\conone_0) &\to \funone_n(\conone_0,\ldots,\conone_0)\\
\funone_n(\conone_1(\varone)) &\to \underbrace{\conone_1(...(\conone_1(\funone_n(\varone)...)}_{n \  times} 
\end{align*}
in order to force the following interpretation $\interp{\funone_n}(X)=n \times X$.
In the same spirit we can encode addition by:
\begin{align*}
\id(\conone_2(\varone,\vartwo)) &\to \add(\varone,\vartwo)\\
\add(\conone_1(\varone),\conone_1(\vartwo)) &\to \conone_1(\conone_1(\add(\varone,\vartwo)))
\end{align*}
in order to force $\interp{\add}(X,Y)=X+Y$ and we can encode multiplication by:
\begin{align*}
\funone_n(\varone) &\to \mult(\varone,\underline{n})\\
\funone_n(\varone) &\to \mult(\underline{n},\varone)\\
\mult(\conone_1(\varone),\vartwo) &\to \add(\vartwo,\mult(\varone,\vartwo))
\end{align*}
in order to force $\interp{\mult}(X,Y)=X \times Y$. We let the reader check that this reasoning can be generalized to any degree. Finally, if $\funone$ is the symbol whose interpretation has been forced to encode the polynomial $P^2$, we add the rule:
$$\funone(\many{\varone}{n}) \to \conone_1(\conone_0) $$ to encode the inequality $P^2 \geq 1$. Finally, let us remark that the same (but more technical) kind of encoding can be performed over $\bQ$.
\end{proof}
Since the encoding presented in the proof of previous Theorem does not depend on the use of a max operator we obtain the following corollary:
\begin{corollary}
The QI synthesis problem is undecidable over $\bN[\overline{X}]$ and $\bQ[\overline{X}]$.
\end{corollary}

\subsection{Decidable synthesis over Max-Poly$\{\bR\}$}
\label{sec:synthesis}
In order to get a precise upper bound, we define two notions of degree. The first notion, called $\times$-degree, corresponds to the maximal power of a polynomial whereas the second notion, called $\emph{\max}$-degree, corresponds to the maximal arity of the $\emph{\max}$ function.
\begin{definition}[Degrees]
Given a function\footnote{The polynomial $0$ will have degrees equal to $0$.}  $Q\neq 0\in \MaxPoly\{\bK\}$  of arity $n$ and normal form $\max (P_1, \ldots, P_k)$, with $P_i$ polynomials, then the $\max$-\emph{degree} of $Q$ is equal to $k$.\\
Moreover,  if $P_i$ is a polynomial of degree $d_i$, where the degree of a $n$-ary polynomial of the shape $\sum_{l=1}^k \alpha_l X_{1}^{i^l_{1}}X_{2}^{i^l_{2}}...X_{n}^{i^l_{n}}$, with $\forall l \in [1,k], \alpha_l \neq 0$,  is equal to $\max_{l \in [1,k]}(\sum^{n}_{j=1}i_{j}^l)$,
then the $\times$-\emph{degree} is equal to $\max_{i  \in [1,k]}d_i$.
\end{definition}
These notions of degree are extended to assignments, the degree of an assignment being the maximal degree of a polynomial in its image.
\begin{definition}
The assignment $\interp{-} \in \MaxPoly\{\bK\}$ is in $\kdMaxPoly\{\bK\}$ if its $\times$-degree and its $\max$-degree are respectively bounded by the constants $\mathsf{d}$ and $\mathsf{k}$. 
\end{definition}
Given an assignment $\interp{-} \in \kdMaxPoly\{\bR\}$ and a function symbol $\funone$ of arity $n$ such that $\funone$ is in the definition domain of $\interp{-}$. By Proposition~\ref{prop:norma}, the assignment of $\funone$ can be written as follows:
$$\interp{\funone}(\overline{X}) = \emph{\max} (P[\funone,1](\overline{X}), \ldots, P[\funone,\mathsf{k}](\overline{X}))$$
where $\overline{X}=X_1, \ldots, X_n$ and $P[\funone,i]$ are polynomials of degree at most $\mathsf{d}$. In other words:
$$P[\funone,i](\overline{X}) = \sum a[\funone, i, j_1, \ldots, j_n] X_1^{j_1} \times \cdots \times X_n^{j_n}$$
with $1\leq i \leq\mathsf{k}$ and $\sum _{\ell=1}^n j_\ell \leq \mathsf{d}$ and where the variable $a[\funone, i, j_1, \ldots, j_n] \in \bR$.\\
Now we show some intermediate lemmata:
\begin{lemma}[Subterm encoding]\label{def:sub} 
Given $\funone$ of arity $n$ and an assignment $\interp{-} \in \kdMaxPoly\{\bR\}$ such that $\interp{\funone}$ is defined, the subterm property can be encoded by the following first order formula:
$$S[\funone]= \bigwedge_{j\in [1,n]} S[\funon,j]$$
with $S[\funone,j] = \forall X_1, \ldots, X_n,\underset{{i\in [1,\mathsf{k}]}}{\bigvee} P[\funone,i](\overline{X})\geq X_j$.\\
In other words, $S[\funone]$ if and only if $\interp{\funone}$ is subterm.
\end{lemma}
\begin{proof}
$\interp{\funone}$ is subterm iff $\forall \many{X}{n},\ \interp{\funone}(\many{X}{n}) \geq \emph{\max}(\many{X}{n})$ iff $\emph{\max} (P[\funone,1](\overline{X}), \ldots, P[\funone,\mathsf{k}](\overline{X})) \geq \emph{\max}(\many{X}{n})$ which is equivalent to $S[\funone]$, by Proposition~\ref{bip}.
\end{proof}
\begin{lemma}[Monotonicity encoding]\label{def:wea}
Given $\funone \in \Fct$ of arity $n$ and an assignment $\interp{-} \in \kdMaxPoly\{\bR\}$ such that $\interp{\funone}$ is defined,  the monotonicity property can be encoded by the following first order formula:
\begin{align*}
M[\funone] &= \forall X_1, \ldots, X_n,\forall Y_1,\ldots,Y_n,\\  
&\underset{{l \in [1,n]}}{\bigwedge}X_l \geq Y_l \implies \underset{{j\in[1,\mathsf{k}]}}{\bigwedge}\underset{{i\in [1,\mathsf{k}]}}{\bigvee} P[\funone,i] (\overline{X}) \geq P[\funon,j] (\overline{Y})
\end{align*}
In other words, $M[\funone]$ if and only if $\interp{\funone}$ is monotonic.
\end{lemma}
\begin{proof}
The proof is just an application of Proposition~\ref{bip}.
\end{proof}
Now we relate the degrees of an expression interpretation with respect to the degree of its symbol interpretations. The main reason for doing so is that we need to encode expression interpretations and not only symbol interpretation in order to encode the rewrite rules of a program.
\begin{proposition}\label{prop:degrees}
Given $\interp{-} \in  \kdMaxPoly\{\bR\}$ and a term $t$, we have $\interp{t} \in\kdmMaxPoly\{\bR\}$.
\end{proposition}
\begin{proof}
By induction on the size of a term $t$.
\end{proof}
Proposition~\ref{prop:degrees} shows that polynomials can be extended to terms. We write: $$\interp{t}(\overline{X}) = \emph{\max}(P[t,1](\overline{X}), \ldots, P[t,\mathsf{k}'](\overline{X}))$$ for some $\mathsf{k}' \leq \mathsf{k}^{\taille{t}}$ and with $P[t,j]$ polynomials of degree bounded by $\mathsf{d}^{\taille{t}}$, whenever the considered assignment is of $\emph{\max}$-degree $\mathsf{k}$ and $\times$-degree $\mathsf{d}$. 

\begin{lemma}[Rule encoding]\label{def:qi}
Given a TRS $\prog$ and an assignment $\interp{-} \in \kdMaxPoly\{\bR\}$, for each rule $l \to_\Regles r$, each inequality can be encoded by:
\begin{align*}
R[l\to r] & = \forall X_1, \ldots X_n,\ \bigwedge_{j\in[1,l]}\bigvee_{i\in[1,n]} P[l,i](\overline{X}) \geq P[r,j](\overline{X})
\end{align*}
with $n\leq \mathsf{k}^{\taille{l}}$ and $l \leq \mathsf{k}^{\taille{r}}$. \\
In other words, $R[l\to r]$ if and only if $\interp{l} \geq \interp{r}$ is satisfied.
\end{lemma}
\begin{proof}
By combining Propositions~\ref{bip} and~\ref{prop:degrees}.
\end{proof}

\begin{proposition}[QI encoding]\label{QIE}
Given a TRS $\prog$, whose symbols have maximal arity $n$, define the first order formula: $$QI[\prog] = \exists a[\funone,i,j_1, \ldots, j_n] \in \bR,  (\bigwedge_{\funtw \in \Functions} (S[\funtw] \wedge M[\funtw])) \wedge  (\bigwedge_{l \to_{\Regles} r \in \Regles } R[l\to r])$$
$QI[\prog]$ is true if and only if there is an assignment $\interp{-}$ that is a quasi-interpretation of $\prog$.
\end{proposition}
\begin{proof}
All the properties of QI are satisfied by  Lemmata~\ref{def:sub}, \ref{def:wea}, and \ref{def:qi}
\end{proof}

\begin{theorem}\label{thm:synthese2}
$\forall \mathsf{k},\mathsf{d} \in \bN$, the QI synthesis problem is decidable in exponential time (in the size of the program) over $\kdMaxPoly\{\bR\}$.
\end{theorem}
\begin{proof}
Given a TRS $\prog$, by Proposition~\ref{QIE}, the QI synthesis problem can be turned into checking the satisfaction of the formula $QI[\prog]$. Note that we can extrude all the quantifiers of the formula $QI[\prog]$, after a careful $\alpha$-conversion, obtaining a new formula under prenex normal form with only one alternation between a block of existential quantifiers (encoding the polynomials multiplicative coefficients) and one block of universal quantifiers (encoding program variables) and we apply the same reasoning than in Theorem~\ref{thm:synthese} (using Theorem~\ref{roy} again). Note that the exponential upper bound lies in the fact that there are only two blocks of quantifiers ($m=2$). 
\end{proof}

\begin{corollary}
The QI synthesis problem is decidable in exponential time over $\bR[\overline{X}]$.
\end{corollary}

\subsection{Another interest in the use of reals}
The interest of considering quasi-interpretations over the reals does not only rely on the decidability result of Theorem~\ref{thm:synthese2}. Indeed, we have an analog result to Theorem~\ref{Luc} over $\MaxPoly$ quasi-interpretations. It states that there exist programs that do not have any quasi-interpretation over ${\MaxPoly}\{\bQ\}$ and, a fortiori ${\MaxPoly}\{\bN\}$, but that admit a quasi-interpretation over ${\MaxPoly}\{\bR\}$.
\begin{theorem}\label{RpasQ}
There are TRS having a quasi-interpre\-tation over $\MaxPoly\{\bR\}$, whereas they do not have any quasi-interpretation over $\MaxPoly\{\bQ\}$.
\end{theorem}
\begin{proof}
We build such a TRS in order to enforce its quasi-interpretation $\interp{-} \in \MaxPoly\left\{\bR\right\}$ to have an irrational coefficient. Our proof is based on additive QI but we claim that there is a similar proof for the general case. The existence of an infinite number of such TRS follows since we can add infinitely many rules with fresh function symbols on such a program. Moreover we may add the following rule:
$$ \id(x)\to \id(\id(x))$$
It enforces the function symbol $\id$ to have a quasi-interpreta\-tion of the shape $\interp{\id}(X)=X$ (otherwise if $\interp{\id}(X)>X$, we have $\interp{\id}(\interp{\id}(X))>\interp{\id}(X)$ and there is no QI for such a program). 
Consider a fresh $2$-ary function symbol $\funtwo$, a $0$-ary constructor symbol $\emph{\zero}$ and  a $2$-ary constructor symbol $\conone$ such that: $\interp{\emph{\zero}}=0$ and $\interp{\conone}(X,Y)=X+Y+1$. Consider the following rule:
$$ \id(\emph{\zero}) \to \funtwo(\emph{\zero},\emph{\zero})$$
If $\interp{-}$ is a quasi-interpretation of the TRS then the following inequality holds:
$$ 0\geq\interp{\funtwo}(0,0)$$
Now consider adding the rule:
\begin{align*}
\id(\conone(\conone(y,y),\conone(y,y))) &\to \funtwo(\conone(\zero,\zero),y)
\end{align*}
$\interp{-}$ has to satisfy that:
\begin{align*}
4 \times Y+3 &\geq \interp{\funtwo}(1,Y)
\end{align*}
Consequently, $\interp{\funtwo}(X,Y)$ has a $\times$-degree at most $1$ in $Y$. Otherwise, for an arbitrary large $Y$, the above inequality is no longer satisfied. Consequently, there is a set $I$ of indexes and polynomials $R_i$ and $S_i$ such that $\interp{\funtwo}(X,Y)=\emph{\max}_{i \in I}(R_i(X)\times Y+S_i(X))$ and $\interp{\funtwo}(0,0)=\emph{\max}_{i \in I}(S_i(0))=0$.\\
Now consider two $1$-ary fresh constructor symbols $\cona$ and $\conb$ such that $\interp{\cona}(X)=X+k$ and $\interp{\conb}(X)=X+k'$, for some $k,k' \in \bN$. Finally, add the following rules:
\begin{align*}
\id(\conb(\conb(\emph{\zero}))) &\to \funtwo(\emph{\zero},\funtwo(\emph{\zero},\conb(\emph{\zero})))\\
\id(\conb(\emph{\zero})) &\to \funtwo(\emph{\zero},\cona(\emph{\zero}))\\
\funtwo(\emph{\zero},\cona(\emph{\zero})) &\to \conb(\emph{\zero})\\
\funtwo(\emph{\zero},\conb(\emph{\zero})) &\to \cona(\cona(\emph{\zero}))
\end{align*}
All these rules correspond to the following inequalities:
\begin{align*}
2 \times k' &\geq \emph{\max}_{i \in I}(R_i(0))^2 \times k'\\
k' &\geq \emph{\max}_{i \in I}(R_i(0)) \times k \\
\emph{\max}_{i \in I}(R_i(0)) \times k &\geq k'\\
\emph{\max}_{i \in I}(R_i(0)) \times k' &\geq 2 \times k
\end{align*}
The first inequality guarantees that $2 \geq \emph{\max}_{i \in I}(R_i(0))^2$ since $k' \geq 1$. We deduce from second and third inequalities that $k'  = \emph{\max}_{i \in I}(R_i(0)) \times k$. Substituting $\emph{\max}_{i \in I}(R_i(0)) \times k$ to  $k'$ in the last inequality, we obtain $\emph{\max}_{i \in I}(R_i(0))^2 \times k \geq 2 \times k$ and, consequently, $\emph{\max}_{i \in I}(R_i(0))^2 \geq 2$, since $k \geq 1$. Finally, $\emph{\max}_{i \in I}(R_i(0))=\sqrt{2}$ and the program only admits irrational quasi-interpretations. In particular, it admits the following quasi-interpretation: $\interp{\emph{\zero}}=0$, $\interp{\emph{\cona}}(X)=X+1$,
$\interp{\emph{\conb}}(X)=X+\sqrt{2}$, $\interp{\emph{\conone}}(X,Y)=X+Y+1$, $\interp{\id}(X)=X$ and $\interp{\funtwo}(X,Y)= \emph{\max}(\sqrt{2} (X+1)Y, X, Y)$.
\end{proof}

\subsection{The QI synthesis problem over $\MaxPlus$}
\subsubsection{$\NP$-hardness results}
The complexity of the QI synthesis problem over $\MaxPoly$ encourage us to consider smaller function sets. In this perspective, Amadio~\cite{Amadio03} has considered assignments in $\MaxPlus\left\{\bN \right\}$\footnote{Indeed Amadio considers polynomials with variables and additive coefficients over $\bQ$ but with multiplicative coefficients over $\bN$, consequently restricting the shape of allowed interpretations, whereas we will explicitly consider all coefficients in $\bQ$ when referring to $\MaxPlus$ $\left\{\bQ\right\}$. Also note that real numbers are not considered in Amadio's result.}. He has demonstrated that the QI synthesis problem is still a hard problem even on such a small set of functions.
\begin{definition}
Let ${\MaxPlus}\left\{\bK\right\}$ be the set of functions obtained using constants and variables ranging over $\bK$ and arbitrary compositions of the operators $+$ and $\emph{\max}$. 
\end{definition}
Now we state a normalization result that is just a corollary of Proposition~\ref{prop:norma}:
\begin{proposition}
 Each function $Q \in {\MaxPlus}\{\bK\}, Q \neq 0$, can be written into the following normal form:
\begin{align*}
  Q(X_1,\ldots, X_n) & =\max_{i\in I}(\sum^{n}_{j=1}\alpha_{i,j}X_{j}+a_{i})
\end{align*}
for some finite set of indexes $I \subset \bN$ and coefficients $\alpha_{i,j}, a_i \in \bK$, $\forall i \in I, j \in [1,n]$.
\end{proposition}

\begin{theorem}[Amadio~\cite{Amadio03}]\label{RA01}
The additive QI synthesis problem is $\NP$-hard over ${\MaxPlus}\left\{\bN\right\}$.
\end{theorem}
In what follows, we will show that the QI synthesis problem remains $\NP$-hard over $\MaxPlus\left\{\bR\right\}$. One could have expected a better result by a naive analogy with linear programming that is $\Ptime$-complete over $\bR$ and $\NP$-complete over $\bN$. This result is inspired by the $\NP$-hardness proof suggested in Amadio~\cite{Amadio03}. However since the quasi-interpretation coefficients are ranging over $\bR$ instead of $\bN$ it generates some technical encoding problems. Indeed, properties of the shape ``If $x + y = 1$ then either $x=1$ and $y=0$ or the converse" hold over $\bN$ but not over $\bR$. More constraints are thus needed on the considered TRS to encode a reduction from a $\NP$-complete problem.
\begin{theorem}\label{thm:synthese3}
The additive quasi-interpretation synthesis problem is $\NP$-hard over $\MaxPlus\left\{\bR \right\}$.
\end{theorem}
The complete proof with key-ingredients is in the Subsection~\ref{nphard}. 
It proceeds by reducing a 3-CNF problem into a synthesis problem for $\MaxPlus\left\{\bR \right\}$. 
The reduction follows Amadio~\cite{Amadio03}. The main difference is that the property  $\sum^{n}_{j=1}\alpha_{i,j}=1 \Rightarrow (\exists j$ such that $\alpha_{i,j}=1$ and $\forall k \neq j$ $ \alpha_{i,k}=0)$ holds over $\bN$ but
no longer holds over reals or rationals. We overcome this problem by adding new rules that give sufficient constraints on the considered assignments to allow us to recover such a property. 
The end of our proof follows Amadio's proof that encodes literals into a synthesis problem: a function symbol $\funone_i$ having a quasi-interpretation $\interp{\funone_i}=\alpha_{1}X_{1}+\alpha_{2}X_{2}$ satisfying ($\alpha_{1}=1$ and $\alpha_{2}=2$) or ($\alpha_{1}=2$ and $\alpha_{2}=1$) is associated to each literal $\varone_i$ of a 3-CNF formula $\phi$. We suppose that some fixed constant $k \geq 1$ (respectively $2k$) is the additive constant corresponding to the interpretation of a constructor symbol $\conone$ and is an encoding of the truth value $\true$ (resp.~$\false$). If the first literal of a disjunction $D$ in $\phi$ is $\varone_{i}$, we associate inputs $(\conone(\zero),\zero)$ to the function symbol $\funone_{i}$. In this case, we have $\interp{\funone_{i}(\conone(\zero),\zero)}=\alpha_{1}\times k$ and $\interp{\funone_{i}}$ will correspond to $\true$ if and only if $\alpha_{1}=1$, that is $\interp{\funone_{i}}(X_1,X_2)=X_{1} + 2\times X_{2}$. If the first literal of $D$ is $\neg\varone_{i}$, we associate inputs $(\zero, \conone(\zero))$ to the function symbol $\funone_{i}$. In this case, we have $\interp{\funone_{i}(\conone(\zero),\zero)}=\alpha_{2}\times k$ and $\interp{\funone_{i}}$ will correspond to $\true$ if and only if $\alpha_{2}=1$, that is $\interp{\funone_{i}}(X_1,X_2)=2\times X_{1} + X_{2}$. Finally we require, using constraints (generated by fresh rules) on the QI, that at least one literal (or its negation) is evaluated to $k$ in each disjunction of $\phi$ by requiring that at most 2 literals of each disjunction are evaluated to $\false$. The provided reduction is polynomial in the size of the formula $\phi$.

\begin{corollary}\label{thm:synthese4}
The additive quasi-interpretation synthesis problem is $\NP$-hard over $\MaxPlus\left\{\bQ\right\}$.
\end{corollary}
\begin{proof}
Just notice that the proof presented in Subsection~\ref{nphard} also holds on $\bQ$. 
\end{proof}

\subsubsection{Proof of $\NP$-hardness over $\MaxPlus\{\bR\}$}\label{nphard}
In this section, we show the $\NP$-hardness of the synthesis problem over $\MaxPlus\{\bR\}$ by exhibiting a reduction of every 3-CNF formula satisfiability problem into a quasi-interpretation synthesis problem. For that purpose, we need some intermediate and technical propositions. 
\begin{proposition}\label{prop:qimaxplus}
Given a TRS $\prog$ having a quasi-interpretation $\interp{-} \in \MaxPlus\{\bR\}$. For every $\funone \in \Fct$ such that $\interp{\funone}(\many{X}{n})=\max_{i\in I}(\sum^{n}_{j=1}\alpha_{i,j}\times X_{j}+a_{i})$ we have: 
$$\forall j\leq n,\ \exists i\in I,\ \alpha_{i,j} \geq 1$$
\end{proposition}
\proof{
Suppose that $\exists j \leq n, \forall i \in I,\ \alpha_{i,j} <1$ holds and let $j_0$ be the value of index $j$ on which it holds.\\
Now take the particular values $x_k=0,\ \forall k \neq j_0$ and $x_{j_0}> \emph{\max}_{i \in I}(a_{i}/(1-\alpha_{i,j_0}))$ we have:
\begin{align*}
\interp{\funone}(x_1,\ldots,x_{j_0} ,\ldots,x_n)&= \emph{\max}_{i\in I}(\alpha_{i,j_0}\times x_{j_0}+a_{i}) \\
&< \emph{\max}_{i\in I}(\alpha_{i,j_0}\times x_{j_0}+(1-\alpha_{i,j_0})\times x_{j_0})\\
&<x_{j_0}
\end{align*}
Note that $x_{j_0}$ is clearly defined since  $\forall i \in I,\ \alpha_{i,j_0} <1$. Consequently, this contradicts the subterm property stating that $\forall j \leq n, \ \forall X_j \in \bR, \ \interp{\funone}(\many{X}{n}) \geq X_j$.
\qed}
\begin{proposition}\label{pr:assign:prop}
There exist a TRS $\prog$ and a function symbol $\funone \in \Functions$ such that if $\prog$ has an additive quasi-interpretation $\interp{-} \in \MaxPlus\{\bR\}$ then at least one of the following conditions holds:
\begin{enumerate}
\item $\interp{\funon}(X_{1},\ldots,X_{n})=\max(X_{1},\ldots,X_{n})$
\item $\interp{\funon}(X_{1},\ldots,X_{n}) = \max_{i\in I}(\sum^{n}_{j=1}\alpha_{i,j}\times X_{j})$ (i.e.~$\interp{\funone}(0,\ldots,0)=0$)
\item $\interp{\funon}(X_{1},\ldots,X_{n})  = \sum^{n}_{j=1}\alpha_{i,j}\times X_{j}$
\end{enumerate}
\end{proposition}
\begin{proof}
\begin{enumerate}
\item We show the first equality by generating the rules of $\Regles$ in order to constraint the quasi-interpretation of $\funone$. Suppose that $\funone$ admits a quasi-interpretation of the shape $ \interp{\funone}(\many{X}{n})=\emph{\max}_{i\in I'}(\sum^{n}_{j=1}\alpha_{i,j}\times X_{j}+a_i)$ and consider adding the following rule:
$$ \funone(\varone_{1},\ldots,\varone_{n}) \to \funone(\funone(\varone_{1},\ldots,\varone_{n}), \ldots, \funone(\varone_{1},\ldots,\varone_{n}))$$
If $\interp{-}$ is a quasi-interpretation then it has to satisfy: $$\interp{\funone(\varone_{1},\ldots,\varone_{n}) }\geq \emph{\max}_{i\in I'}((\sum^{n}_{j=1}\alpha_{i,j}) \times \interp{\funone(\varone_{1},\ldots,\varone_{n}) }+a_{i})$$ Consequently, $\forall i \in I'$, $\sum_{j = 1}^n \alpha_{i,j} \leq 1$. Using Proposition~\ref{prop:qimaxplus}, we have that for each $j$ there is a particular $i_j \in I'$ such that $\alpha_{i_j,j}\geq 1$. Combined with previous inequality, it implies that $\alpha_{i_j,j}=1$ and $\forall l,\  l\neq j,\alpha_{i_j,l}=0$.\\
So we can write the quasi-interpretation of $\funone$ as follows: $$\interp{\funone}(\many{X}{n})=\emph{\max}(X_{1}+a_{i_1},\ldots,X_{n}+a_{i_n},\emph{\max}_{i \in I}(\sum^{n}_{j=1}\alpha_{i,j}\times X_{j}+a_{i}))$$
with $I=I'-\left\{\many{i}{n}\right\}$ and $\forall j \leq n, \forall i \in I, \ \alpha_{i,j} < 1$. \\
For an arbitrarily large value $x \in \bR$ (take $x > \emph{\max}_{i \in I}((a_i - a_{i_1})/(1 -\alpha_{i,1}))$), we have $\interp{\funone}(x,0,\ldots, 0) = x + a_{i_1}$, with $a_{i_1} \geq 0$. Indeed $\forall i \in I, \ \alpha_{i,1} \times x +a_i < x +a_{i_1}$. It implies that:
\begin{align*}
\interp{\funone}(x,0,\ldots, 0) &= x + a_{i_1} \\
&\geq \interp{\funone}(\interp{\funone}(x,0,\ldots,0),\ldots,\interp{\funone}(x,0,\ldots,0))\\
&\geq \interp{\funone}(x + a_{i_1},\ldots,x + a_{i_1})\\
&\geq x + 2 \times a_{i_1}
\end{align*}
Consequently, $a_{i_1}=0$. Since we can perform the same reasoning for each constant $a_{i_k}$, the quasi-interpretation can be written:
$$\interp{\funone}(\many{X}{n})=\emph{\max}(X_{1},\ldots,X_{n},\emph{\max}_{i \in I}(\sum^{n}_{j=1}\alpha_{i,j}\times X_{j}+a_{i}))$$
with $\sum^{n}_{j=1}\alpha_{i,j} \leq 1$. Now consider adding the following rule to the program: $$\funone(\conb(\varone_{1},\zero),\ldots \conb(\varone_{n},\zero)) \to \funone(\conb(\varone_{1},\funone(\zero,\ldots\zero)),\ldots \conb(\varone_{n},\funone(\zero,\ldots,\zero)))$$
with  $\conb$ a constructor symbol such that $\interp{\conb}(X,Y)=X+Y+k_\conb$, $k_\conb\geq 1$. In order for $\interp{-}$ to be a QI, it is necessary to check that $$\interp{\funone(\conb(\varone_{1},\zero),\ldots \conb(\varone_{n},\zero))} \geq \interp{\funone(\conb(\varone_{1},\funone(\zero,\ldots\zero)),\ldots \conb(\varone_{n},\funone(\zero,\ldots,\zero)))}$$ It implies by choosing the particular values $\interp{\varone_1}=\ldots=\interp{\varone_n}=x \in \bR$:
{\small $$
\emph{\max}_{i\in I}((\sum^{n}_{j=1}\alpha_{i,j})\times (x + k_{\conb})+a_{i})
\geq \emph{\max}_{i\in I}((\sum^{n}_{j=1}\alpha_{i,j})\times (x + k_{\conb}+\emph{\max}_{k\in I}(a_{k}))+a_{i})
$$}
Suppose that $l$ is the index for which $\emph{\max}_{i \in I}(\sum^{n}_{j=1}\alpha_{i,j})$ is reached. For an arbitrary large $x$ and since $\sum^{n}_{j=1}\alpha_{l,j}=1$ and  $a_{l}=0$, we have $x+k_{\conb}\geq x+k_{\conb}+\emph{\max}_{k\in I}(a_{k})$. It implies $a_{k}=0,\ \forall k \in I$.
Finally we have:
$$\interp{\funone}(\many{X}{n})=\emph{\max}(\emph{\max}(X_{1},\ldots,X_{n}),\emph{\max}_{i \in I}(\sum^{n}_{j=1}\alpha_{i,j} \times X_{j}))
$$
with $\sum^{n}_{j=1}\alpha_{i,j}\leq 1$. Since $\forall X_{1},\ldots,\forall X_{n} \in \bRp,\ \emph{\max}(X_{1},\ldots,X_{n})\geq \emph{\max}_{i \in I}(\sum^{n}_{j=1}\alpha_{i,j} \times X_{j})$ holds, we obtain: 
$$\interp{\funone}(\many{X}{n})=\emph{\max}(X_{1},\ldots,X_{n})$$
\item Now we show the second equality. Given $\funtwo$ a function symbol such that $\interp{\funtwo}(\many{X}{n})=\emph{\max}_{i \in I}(\sum^{n}_{j=1}\alpha_{i,j} \times X_{j}+a_i)$. We add the rule:
$$\id( \cond(\varone_{1},\ldots,\varone_{n})) \to \cond(\funtwo(\zero,\ldots,\zero),\zero,\ldots,\zero)$$
with $\id$ a function symbol such that $\interp{\id}(X)=X$ (There exists such a function symbol by Proposition~\ref{pr:assign:prop}, item (1)) and $\cond$ a $n$-ary constructor symbol such that $\interp{\cond}(\many{X}{n})=\sum_{i=1}^nX_i +k_\cond$, $k_\cond \geq 1$.
The corresponding assignment has to satisfy $k_{\cond}+\sum^{n}_{j=1}X_{j}\geq k_{\cond}+\emph{\max}_{i\in I}(a_{i})$. It implies that  $\forall i \in I,\ a_{i}=0$. Consequently, $\interp{\funtwo}(\many{X}{n})=\emph{\max}_{i\in I}(\sum^{n}_{j=1}\alpha_{i,j}X_{j})$.
\item Consider a function symbol $\funone$ of arity $n$. Its quasi-interpretation can be constrained to be of the shape:
 \begin{align*}
\interp{\funone}(\many{X}{n}) &= \alpha_{1}\times X_{1}+\cdots+\alpha_{n}\times X_{n}
 \end{align*}
by adding the following rules to the program:
$$ \id(\conone(\varone))\to \conone(\funone(\zero,\ldots,\zero,\varone,\zero,\ldots\zero))$$
with $\varone$ appearing at the i-th position in the right hand side of the rule, $\forall i \in [1,n]$, $\conone$ a 1-ary constructor symbol such that $\interp{\conone}(X)=X+k, \ k \geq 1$, and with $\id$ a function symbol such that $\interp{\id}(X)=X$.
\end{enumerate}
\end{proof}

\begin{proposition}\label{tartanpion}
There exist a TRS $\prog$ and a function symbol $\funone \in \Functions$ of arity $2$ such that if $\prog$ has an additive quasi-interpretation $\interp{-} \in \MaxPlus\{\bR\}$ then the following conditions both hold:
\begin{itemize}
\item $\interp{\funone}(X_1,X_2)=\alpha_{1}\times X_{1}+\alpha_{2}\times X_{2} $
\item $(\alpha_{1}=1\wedge \alpha_{2}=2)\vee (\alpha_{1}=2\wedge \alpha_{2}=1)$
\end{itemize}
\end{proposition}

\begin{proof}
By Proposition~\ref{pr:assign:prop}, we can enforce a $2$-ary function symbol $\funone$  to have the following quasi-interpretation by adding arbitrary rules to constraint its quasi-interpretation:
\begin{center}
$\interp{\funone}(X_1,X_2)=\emph{\max}_{i\in I}(\alpha_{i,1}\times X_{1}+\alpha_{i,2}\times X_{2})$ 
\end{center}
We define $\alpha_{j}=\emph{\max}_{i\in I}(\alpha_{i,j})$, for $j \in \left\{1,2\right\}$, and $\alpha=\emph{\max}_{i\in I}(\alpha_{i,1}+\alpha_{i,2})$. These constants satisfy the following inequality $\alpha_{1}+\alpha_{2}\geq \alpha$.
Now add the following rule to the considered TRS:
$$
\funone(\conb(\varone_{1},\zero),\conb(\varone_{2},\zero)) \to \conb(\funone(\varone_{1},\zero),\funone(\zero,\varone_{1}))
$$
with $\conb$ a 2-ary constructor symbol such that $\interp{\conb}(X)=X+k$ and $\interp{0}=0$. For the particular values $\interp{\varone_{1}}=\interp{\varone_{2}}=x \in \bR$, the corresponding quasi-interpretation has to satisfy $\alpha  \times (x+k)\geq k+(\alpha_{1}+\alpha_{2})\times  x$. Consequently, for an arbitrarily large $x$, $\alpha=\alpha_{1}+\alpha_{2}$ and we can write:
\begin{align*}
\interp{\funone}(X_1,X_2)=\alpha_{1}\times X_{1}+\alpha_{2}\times X_{2}
\end{align*}
since $\exists j \in I,\ \alpha_{j,1}= \alpha_{1}$ and $\alpha_{j,2}=\alpha_{2}$. Indeed it implies that
$\forall X_{1},X_{2}\in \bRp,\ \forall i \in I\, i \neq j,\  \alpha_{j,1} \times  X_{1}+\alpha_{j,2}\times X_{2}\geq\alpha_{i,1} \times X_{1}+\alpha_{i,2} \times X_{2}$. \\
By virtue of the subterm condition, $\alpha_{1},\alpha_{2}\geq 1$. We add new rules over $\funone$ in order to constraint $\alpha_{1}$ and $\alpha_{2}$ to satisfy the following condition $(\alpha_{1}=1\wedge \alpha_{2}=2)\vee (\alpha_{1}=2\wedge \alpha_{2}=1)$:
\begin{align*}
\funone(\conone(\varone_{1}),\conone(\varone_{2})) &\to \conone(\conone(\conone(\zero))) \\
\id(\conone(\conone(\conone(\varone)))) &\to \funone(\conone(\zero),\conone(\zero))
\end{align*}
If $\interp{-}$ is a quasi-interpretation, $\conone$ is a 1-ary constructor symbol such that $\interp{\conone}(X)=X+k$ and $\id$ is a function symbol such that $\interp{\id}(X)=X$ (such a symbol exists by Proposition~\ref{pr:assign:prop}), we deduce from these rules that $\alpha_{1}+\alpha_{2}= 3$. By adding the rule:
\begin{align*}
\id(\conone(\conone(\varone))) \to \funone(\funone(\zero,\conone(\zero)),\zero)
\end{align*}
we check that $2\times k+x\geq \alpha_{1}\times \alpha_{2}\times k$. In other words, $2\geq \alpha_{1}\times \alpha_{2}$.  Since $\alpha_{1}=3-\alpha_{2}$, we have to check the inequality $\alpha_{1}^{2}-3\times \alpha_{1}+2\geq0$ with $2\geq\alpha_{1}\geq 1$. The only corresponding solutions are ($\alpha_{1}=1\wedge \alpha_{2}=2)\vee (\alpha_{1}=2\wedge \alpha_{2}=1$). 
\end{proof}

\setcounter{theorem}{18}
\begin{theorem}
The additive quasi-interpretation synthesis problem is $\NP$-hard over $\MaxPlus\left\{\bR \right\}$.
\end{theorem}
\begin{proof}
We encode the satisfiability of a 3-$\mathsf{SAT}$ problem under 3-CNF into a QI synthesis problem. Given a 3-CNF formula $\phi$, we generate a TRS $\prog$ such that $\prog$ admits a quasi-interpretation if and only if $\phi$ is satisfiable. In this perspective, we associate to each literal $\varone_{i}$ appearing in a given 3-CNF formula $\phi$, a fresh 2-ary function symbol $\funone_{i}$ and it corresponding rules such that $\interp{\funone_{i}}(X_1,X_2)= \alpha^{i}_{1}\times X_{1}+ \alpha^{i}_{2}\times X_{2}$, with $(\alpha^i_{1}=1\wedge \alpha^i_{2}=2)\vee (\alpha^i_{1}=2\wedge \alpha^i_{2}=1)$. Note that this is made possible by Proposition~\ref{tartanpion}.\\
 The table of Figure~\ref{tableau} subsumes the distinct values taken by the quasi-interpreta\-tion $\interp{\funone_{i}}$ wrt its coefficients and its inputs, for some constructor symbols $\conone$ and $\zero$ such that $\interp{\conone}(X)=X+k$ and $\interp{\zero}=0$.
\begin{figure}[!h]
\begin{gather*}
\begin{tabular}{|c|c|c|}
\hline
Coefficients of $\interp{\funone_{i}}$: & inputs: & Value of:
\\ 
$(\alpha^{i}_{1},\alpha^{i}_{2})$ &$(\varone_{1},\varone_{2})$ &$ \interp{\funone_i(\varone_1,\varone_2)}$
                   \\[2pt]
\hline 
(1,2)  &  $(\conone(\zero),\zero)$&              $k$  \\
(1,2)  &  $(\zero,\conone(\zero))$&              $2\times k$ \\
(2,1)  &  $(\conone(\zero),\zero)$&              $2\times k$ \\
(2,1)  &  $(\zero,\conone(\zero))$&              $k$  \\[2pt]
\hline
\end{tabular}
\end{gather*}
\caption{Values of $ \interp{\funone_{i}}$ wrt its coefficients and inputs}
\label{tableau}
\end{figure}

Let the constant $k$ (respectively $2\times k$) encode the truth value $\true$ (respectively $\false$). If a literal corresponds to $\true$ (resp.~$\false$) then we will encode this information by constraining $\funone_{i}$ to have a quasi-interpretation equal to $X_{1} + 2\times X_{2}$ (resp.~$2\times X_{1} + X_{2}$). \\
Given a disjunction $D$ of the formula  $\phi$, there are two possibilities:
\begin{itemize}
\item[(i)] If the first literal of $D$ is $\varone_{i}$, we associate inputs $(\conone(\zero),\zero)$ to the function symbol $\funone_{i}$. In this case, we have $\interp{\funone_{i}(\conone(\zero),\zero)}=\alpha^i_{1}\times k$ and $\interp{\funone_{i}}$ will correspond to $\true$ if and only if $\alpha^i_{1}=1$, that is $\interp{\funone_{i}}(X_1,X_2)=X_{1} + 2\times X_{2}$. 
\item[(ii)] If the first literal of $D$ is $\neg\varone_{i}$, we associate inputs $(\zero, \conone(\zero))$ to the function symbol $\funone_{i}$. In this case, we have $\interp{\funone_{i}(\conone(\zero),\zero)}=\alpha^i_{2}\times k$ and $\interp{\funone_{i}}$ will correspond to $\true$ if and only if $\alpha^i_{2}=1$, that is $\interp{\funone_{i}}(X_1,X_2)=2\times X_{1} + X_{2}$.
\end{itemize}
Using the notation $\phi_D^\varone$ to represent the arguments of the function symbol encoding $\varone$ in the disjunction $D$, we have:
 \begin{center}
 $\phi_D^\varone= \left\{ \begin{array}{c} \conone(\zero),\zero\ \text{if } \varone \text{ appears in } D \ \\ \zero,\conone(\zero)\ \text{if } \neg \varone \text{ appears in } D \end{array} \right.$
  \end{center}
  $\interp{\funone(\phi_D^\varone)}$ is equal to $k$ if ($\interp{\funone}$ corresponds to $\true$ and $\varone$ appears in $D$) or ($\interp{\funone}$ corresponds to $\false$ and $\neg \varone$ appears in $D$).\\
  $\interp{\funone(\phi_D^\varone)}$ is equal to $2\times k$ if ($\interp{\funone}$ corresponds to $\true$ and $\neg \varone$ appears in $D$) or ($\interp{\funone}$ corresponds to $\false$ and $\varone$ appears in $D$).\\
It remains to encode disjunctions: To each disjunction $D$ in the formula $\phi$ and containing literals $\varone_{i}$, $\varone_{j}$ and $\varone_{l}$, we associate the following rule:
 \begin{align*}
 \id(\conone(\conone(\conone(\conone(\conone(\varone)))))) \to \funone(\funone_{i}(\phi_D^{\varone_{i}}),\funone_{j}(\phi_D^{\varone_{j}}),\funone_{l}(\phi_D^{\varone_{l}}))
 \end{align*}
$\funone$ and $\id$ being symbols defined by rewrite rules such that their quasi-interpreta\-tions are defined by $\interp{\id}(X)=X$ and $\interp{\funone}(X_1,X_2,X_3)=\alpha_1 \times X_1 + \alpha_2 \times X_2 + \alpha_3 \times X_3$. Note that such symbols exist by items (1) and (3) of Proposition~\ref{pr:assign:prop}. Moreover,  by Proposition~\ref{prop:qimaxplus},  $\alpha_1,\alpha_2,\alpha_3 \geq 1$.
The quasi-interpretation of the obtained TRS has to satisfy:
 \begin{align*}
5\times k+X &\geq \interp{\funone_{i}(\phi_D(\varone_{i}))}+\interp{\funone_{j}(\phi_D(\varone_{j}))}+\interp{\funone_{l}(\phi_D(\varone_{l}))}
 \end{align*}
This inequality enforces at least one of the $\interp{\funone_{p}(\phi_D(\varone_{p}))}$ (for $p \in \{i,j,l\}$) to have value $k$ (i.e.~to be $\true$) and enforces at most two to have value $2k$. Otherwise it is not satisfied because $\neg (5 \times k \geq 6 \times k)$. We encode in the same spirit all the disjunctions of $\phi$. Every assignment satisfying $\phi$ will clearly correspond to the existence of a suitable quasi-interpretation for the program. Indeed, just take $\interp{\funone_i}(X_1,X_2)=X_1+2\times X_2$ (resp. $2 \times X_1+ X_2$) for each litteral assigned to $\true$ (resp. $\false$). Conversely, if the program admits a quasi-interpretation then every disjunction maybe evaluated to true by assigning the truth value $\true$ to each literal corresponding to a quasi-interpretation of the shape $X_1  + 2 \times X_2$. Finally, we have encoded a 3-CNF problem into a QI synthesis problem over $\MaxPlus\left\{\bRp\right\}$ using a polynomial time reduction. Indeed the number of added rules is linear in the the size the formula since each intermediate proposition only introduce a constant number of new rules in the considered TRS. Consequently, this problem is $\NP$-hard.
\end{proof}

\subsubsection{$\NP$-completeness over $\MaxPlus$}
After studying $\NP$-hardness results over $\MaxPlus$, we are interested in completeness results on this function space. We start to introduce the first result demonstrated by Amadio in~\cite{Amadio03} over $\MaxPlus\left\{0,1\right\}$.
Let $\MaxPlus\left\{0,1\right\}$ be the set of functions obtained using constants ranging over $\{0,1\}$ and variables ranging over $\bQ$ and arbitrary compositions of the operators $+$ and $\emph{\max}$\footnote{Such functions were called multi-linear polynomials in~\cite{Amadio03}.}.
\begin{theorem}[Amadio~\cite{Amadio03}]
The additive QI synthesis problem is $\NP$-complete over $\MaxPlus\left\{0,1\right\}$.
\end{theorem}
We try to extend this result to $\bN$ and $\bQ$. For that purpose, we focus on the \emph{QI verification problem} that consists in checking that an assignment of a given TRS is a  quasi-interpretation. We show that this problem can be solved in polynomial time over $\MaxPlus$ if we consider assignment of $\emph{\max}$-degree $\mathsf{k}$ and $+$-degree $\mathsf{d}$ polynomially bounded by the TRS size. This is not a restrictive condition since most of the TRS admitting a quasi-interpretation in $\MaxPlus$ satisfy it. Indeed arity of the max is indexed by the number of rules in the TRS. Each rule may create a new constraint and may consequently increase the max arity by 1. Finally, the arity of the $+$-degree is trivially indexed by the size of expressions in the rules. 
\begin{definition}[$+$-degree and $\emph{\max}$-degree]
Given a function $Q$ of arity $n$ in $\MaxPlus\{\bK\}$ of normal form $\max(P_1,\ldots,P_m)$ with: $$P_i=\sum_{ j\in [1,n]}\alpha_{i,j} \times X_j + \alpha_{i,0}$$ its $+$-degree is equal to $\max_{ j\in [0,n], i \in [1,m]}\alpha_{i,j}$. In other words, the $+$-degree of $Q$ is its greatest multiplicative coefficient. Its $\max$-degree is equal to $m$.
\end{definition}


\begin{definition}
Let $\kdMaxPlus\{\bK\}$ be the set of $\MaxPlus\{\bK\}$ functions of $+$-degree bounded by the constant $\mathsf{d}$ and $\max$-degree bounded by the constant $\mathsf{k}$.\\
Given a TRS $\prog$ and an assignment $\interp{-}$, $\interp{-} \in \kdMaxPlus\{\bK\}$ if: $$\forall l \to r \in \Regles,\ \interp{l},\interp{r} \in \kdMaxPlus\{\bK\}$$
\end{definition}

\begin{theorem}[Verification]\label{complet}Given a TRS $\prog$ and an assignment $\interp{-} \in \kdMaxPlus\left\{\bRp\right\}$, we can check in polynomial time in $\mathsf{d}$ and $\mathsf{k}$ that $\interp{-}$ is a quasi-interpretation of $\prog$.
\end{theorem}
\begin{proof}
Given a TRS $\prog$ and an assignment $\interp{-}$, for each rule of the shape $l \to r$, we can compute $\interp{l}$ and $\interp{r}$ in polynomial time relatively to $\mathsf{k}$ and $\mathsf{d}$, by definition of $\kdMaxPlus$ assignments. Consequently, it remains to check that the inequalities of the shape $\interp{l} \geq \interp{r}$ are satisfied (we also have to check some inequalities for monotonicity and subterm properties that we omit). The total number of such inequalities is polynomially bounded by the TRS size $r$. Moreover, by Proposition~\ref{bip}, we can eliminate the $\emph{\max}$ operators so that each inequality is transformed into the conjunctions and disjunction of $\mathsf{k}^2$ inequalities of the shape $P \geq Q$, with $P,Q \in \MaxPlus\{\bR\}$.
Such inequalities have size polynomially bounded by $\mathsf{k}$ and $\mathsf{d}$. We can check their satisfaction in polynomial time in these two parameters using linear programming over $\bR$, iterating this procedure at most $r\times \mathsf{k}^2$.
\end{proof}

\begin{theorem}\label{tyty}
The additive quasi-interpretation synthesis problem is $\NP$-com\-plete over $\kdMaxPlus\left\{\bN \right\}$, for $\mathsf{d} \geq 2$.
\end{theorem}
\begin{proof}
The $\NP$-hardness has been demonstrated in Theorem~\ref{RA01}. For that purpose, we need a $+$-degree of at least $2$ in our encoding of $3$-CNF. We have shown in Theorem~\ref{complet}, that the verification problem that consists in checking for a candidate assignment that it is a quasi-interpretation can be solved in polynomial time (if variables are extended to $\bR$). It remains to see that the size of each solution is bounded polynomially by the input size (the TRS size): it is the case since its degrees are bounded by constants $\mathsf{k}$ and $\mathsf{d}$.
\end{proof}

\begin{theorem}\label{tete}
Let $\bQ_{\leq \textbf{\emph{d}}}$ be the subset of $\bQ$ such that every rational has both numerator and denominator bounded by $\mathsf{d}$.
The additive quasi-interpretation synthesis problem is $\NP$-complete over $\kdMaxPlus\left\{\bQ_{\leq \mathsf{d}} \right\}$, for $\mathsf{d} \geq 2$.
\end{theorem}
\begin{proof}
Every rational from $\bQ_{\leq \mathsf{{d}}}$ can be encoded by two integers smaller than $\mathsf{d}$ and, consequently, has a size bounded polynomially by $\mathsf{d}$.
\end{proof}
Such a result does not hold in general for $\bR$ because of the representation problem in such a space: we do not know how to encode the data since a real number is generally not bounded even if we have bounded degrees.\footnote{This is Corrigendum to~\cite{BMMP05} where it was wrongly stated that the QI synthesis problem is $\NP$-complete over ${\kdMaxPlus}\left\{\bR \right\}$.} 

\section{Dependency Pair interpretations}\label{S6}
\subsection{DP-interpretations as sup-interpretations}
The notion of sup-interpretation was introduced in~\cite{MP09} in order to increase the intensionality of interpretation methods. One of the main distinction with quasi-interpretations lies in the subterm property (cf. Definition~\ref{qidef}): sup-interpreta\-tions do not need to satisfy such a property. Consequently, the subterm property drastically restricts the sup-interpretation space. For example, a function defined by $\funone(x,y) \to x$ has a QI at least equal to $\interp{\funone}(X,Y)=\emph{\max}(X,Y)$ whereas one would expect its sup-interpretation to be equal to $\theta(\funone)(X,Y)=X$ since the second parameter is dropped. To overcome this problem, we introduce a new notion of sup-interpretations, namely DP-interpretations, based on the notion of dependency pair (DP) introduced by Arts and Giesl~\cite{AG00} for showing program termination. Note that a similar notion was introduced in~\cite{MP08} for characterizing $\FPtime$ but was not related to the notion of sup-interpretation. A last point to mention is that DP-interpretations are not a DP-method since they do not ensure termination but rather a method for space analysis inspired by DP-methods.
We start by briefly reviewing the notion of dependency pair: 
\begin{definition}[DP]
Given a TRS $\prog$, the set of dependency pair symbols $\Fct^{\sharp}$ is defined by $\Fct^{\sharp}= \Fct \cup \{ \funone^{\sharp} \ | \ \funone \in \Fct \}$, $\funone^{\sharp}$ being a fresh function symbol of the same arity as $\funone$. Given a term $t = \funone(\many{t}{n})$, let $t^{\sharp}$ be a notation for $\funone^{\sharp}(\many{t}{n})$.\\ A dependency pair is a pair $l^{\sharp} \to u^{\sharp}$ if $u^{\sharp}=\funtw^{\sharp}(\many{t}{n})$, for some $\funtw \in \Fct$, and if there is a context $\conta{\magique}$ such that $l \to \conta{u} \in \Regles$ and $u$ is not a proper subterm of $l$. Let $DP(\Regles)$ be the set of all dependency pairs in $\prog$.
\end{definition}
\begin{definition}[DP-interpretation]
Given a TRS $\prog$, a (additive) DP-interpretation (DPI for short) is a monotonic (additive) assignment $\inter{-}$ over $\bK$ extended to $\Fct^{\sharp}$ by $\forall \funone, \in \Fct,\ \inter{\funone^{\sharp}}=\inter {\funone}$ and which satisfies:
\begin{enumerate}
\item $\forall l \to r \in \Regles, \ \inter{l} \geq \inter{r}$
\item $\forall  l^\sharp \to u^\sharp \in DP(\Regles),\ \inter{l^\sharp} \geq \inter{u^\sharp}$
\end{enumerate}
where the DP-interpretation $\inter{-}$ is extended canonically to terms as usual.
\end{definition}
Notice that the main distinction with QI is that the subterm property has been replaced by Condition 2.
We obtain a result similar to Theorem~\ref{qiissi}:
\begin{theorem}
Given a TRS $\prog$ having an additive DP-interpretation $\inter{-}$ then $\inter{-}$ is a sup-interpretation. 
\end{theorem}
Moreover, we can show that every quasi-interpretation is a DP-interpretation.
\begin{theorem}\label{QIDPI}
Given a TRS $\prog$ having a quasi-interpretation $\interp{-}$, $\interp{-}$ is a DP-interpretation.
\end{theorem}
\begin{proof}
By Definition~\ref{qidef}, $\interp{-}$ is a monotonic assignment which satisfies $\forall l \to r \in \Regles, \ \interp{l} \geq \interp{r}$. Now, take $s^\sharp \to t^\sharp \in DP(\Regles)$. By definition, there is a context $\conta{\magique}$ such that $s \to_\Regles \conta{t} \in \Regles$. For each term $t$, $\interp{\conta{t}} \geq \interp{t}$ since $\interp{\conta{t}}$ is obtained by composition of subterm functions (the subterm property is stable by composition). Consequently, $\interp{s^\sharp} = \interp{s} \geq \interp{\conta{t}} \geq \interp{t} = \interp{t^\sharp}$.
\end{proof}
As expected, the converse property does not hold. There are TRS that admit an (additive) DP-interpretation but no (additive) quasi-interpretation, as illustrated by the following example:
\begin{example}
\begin{align*}
&\half(\zero) \to \zero &&\half(1) \to \zero\\
&\half(x+2) \to  \half(x)+1 && \\
&\loga(x+2)  \to  \loga(\half(x+2))+1 &&\loga(1)  \to  0 
\end{align*}
The above TRS has no additive quasi-interpretation since an additive quasi-interpretation such that $\interp{+1}(X)=X+k$, for $k\geq1$, would have to satisfy the following inequalities:
\begin{align*}
\interp{\loga(x+2)}  &\geq \interp{\loga(\half(x+2))+1}\\
&\geq \interp{\loga(\half(x+2))}+k \geq \interp{\loga(x+2)}+k 
\end{align*}
By subterm and monotonicity properties. However, we let the reader check that it admits the following additive DP-interpretation $\inter{0}=1$, $\inter{+1}(X)=X+1$, $\inter{\half}(X)=(X+1)/2$ and $\inter{\loga}(X)=2\times X$.
\end{example}

\subsection{Decidability results over $\bK[\overline{X}]$, $\MaxPoly\{\bK\}$ and $\MaxPlus\{\bK\}$}
In this section, we review the results of the DPI synthesis problem.
\begin{definition}[DPI synthesis problem]
Given a TRS $\prog$, is there an assignment $\inter{-}$ such that $\interp{-}$ is a DP-interpretation of  $\prog$?
\end{definition}

\begin{theorem}
 The DPI synthesis problem is:
\begin{enumerate}
\item \label{a} undecidable over $\MaxPoly\{\bN\}$ and $\MaxPoly\{\bQ\}$
\item \label{b} decidable in exponential time over $\kdMaxPoly\{\bR\}$
\item \label{c} undecidable over $\bN[\overline{X}]$ and $\bQ[\overline{X}]$
\item \label{d} decidable in exponential time over $\bR[\overline{X}]$
\end{enumerate}
\end{theorem}
\begin{proof}
(\ref{a}) is a corollary of Theorem~\ref{QIUN}. The subterm property is withdrawn and replaced by inequalities on dependency pairs. These inequalities do not change the undecidability of the problem. (\ref{b}) is also a corollary of Theorem~\ref{thm:synthese2} using the same reasoning: the encoding of the subterm property is no longer needed and replaced by the encoding of inequalities on DP. Since the number of DP is at most linear in the size of the program, these new inequalities does not impact the complexity of the algorithm.  (\ref{c}) is a consequence of (\ref{a}) and (\ref{d}) is a consequence of (\ref{b}) because polynomials are functions in $\MaxPoly$.
\end{proof}

\subsection{$\NP$-hardness over $\MaxPlus\{ \bK\}$}
Now we show $\NP$-hardness and $\NP$-completeness results:
\begin{theorem}
The additive DPI synthesis problem is:
\begin{enumerate}
\item \label{aa} $\NP$-hard over ${\MaxPlus}\left\{\bK\right\}$, $\bK \in \{\bN,\bQ, \bR\}$
\item \label{aaa}$\NP$-complete over ${\kdMaxPlus}\left\{\bN\right\}$, $\mathsf{d} \geq 2$
\item\label{aaaa} $\NP$-complete over ${\kdMaxPlus}\left\{\bQ_{\leq \mathsf{d}}\right\}$, $\mathsf{d} \geq 2$\footnote{Cf. previous section}
\end{enumerate}
\end{theorem}
\begin{proof}
(\ref{aa}) We use the encoding in the proof of Theorem~\ref{thm:synthese3}. There is just one difficulty to face: By Theorem~\ref{QIDPI} every QI of a given program is a DPI but the converse does not hold. Consequently, it might be easier to find the DPI of a given program than to find its QI, the solution space being greater. Consequently, we have to enforce that each DPI of the reduction is also a QI. This can be done by adding the following rules to the program, $\forall \funone \in \Fct$ of arity $n$ and $\forall i \in \{1,\ldots,n\}$ :
$$\funone(\many{\varone}{n}) \to \varone_i$$
It enforces that the corresponding additive DPI has to satisfy: $$\forall \many{X}{n},\ \inter{\funone}(\many{X}{n}) \geq \emph{\max}(\many{X}{n})$$
Consequently, $\inter{-}$ is DPI then it is a quasi-interpretation and we obtain that the DPI synthesis problem is NP-hard over ${\MaxPlus}\left\{\bK\right\}$, $\bK \in \{\bN,\bQ, \bQ\}$. (\ref{aaa}) is a direct consequence of (\ref{aa}) and Theorem~\ref{tyty} whereas (\ref{aaaa}) is a consequence of (\ref{aa}) and Theorem~\ref{tete}.
\end{proof}
 To conclude, we have found a better notion than the one of quasi-interpretation from an intensional point of view (i.e.~in terms of algorithms) in order to get a sup-interpretation at equal cost from a synthesis point of view.

\section{Runtime complexity}\label{S7}
\subsection{Runtime complexity functions as sup-interpretations}
As previously stated, sup-interpretation is a tool that inherently deals with space consumption in an extensional way. Consequently, it is natural to link this notion with studies on time consumption of TRS. In an analogy with classical complexity theory, one could expect that a TRS running in polynomial time would lead the programmer to get a polynomial upper bound on the size of the computed value.\\
A good candidate for the notion of time complexity of a TRS is the notion of \emph{runtime complexity} function, a function providing an upper bound on the length of the longest derivation with respect to the size of the initial term. Many studies have demonstrated that termination techniques can be used to study the runtime complexity of a given TRS. See~\cite{HL88,MS08,AMS08,W95}, among others. In this subsection, we show that, as expected, bounding the runtime complexity of a TRS, allows us to recover a sup-interpretation. For that purpose, we introduce usual definitions:
\begin{definition}\label{dc}
The \emph{derivational length} of a terminating term $t$ with respect to a rewrite relation $\to_{\Regles}$ is defined by:
$$\mathsf{dl}(t,\to_\Regles)= \max \{ n \in \bN \ | \ \exists s,\ t \to_\Regles^n s\} $$

The \emph{runtime complexity function} with respect to a rewrite relation $\to_{\mathcal{S}}$ on a set of terms $T$ is defined by:
$$\mathsf{rc}(n,T,\to_\mathcal{S})= \max \{\mathsf{dl}(t,\to_\mathcal{S}) \ | \ t \in T \text{ and } \taille{t} \leq n\} $$
The \emph{runtime complexity function} with respect to a TRS $\prog$ is defined by $$\mathsf{rc}_\Regles(n)=\mathsf{rc}(n,T_b,\to_\Regles)$$ where $T_b$ is the set of basic terms of the shape $t=\funone(\many{v}{n})$ with $\funone \in \Fct$ and $\many{v}{n} \in \Consterms$. 
\end{definition}
We start to show an intermediate result that links the size of a term with respect to the length of its derivation. For that purpose, the size of a TRS $\taille{\prog}$ is defined by $\taille{\prog} = \sum_{l \to r \in \Regles} \taille{l}+\taille{r}$.
\begin{lemma}\label{tyf}
Given a TRS $\prog$, for every terms $t,s$ and every $n \in \bN$ such that $t \to_\Regles^n s$ we have:
$$\taille{s} \leq \taille{t} \times \taille{\prog}^n $$
\end{lemma}
\begin{proof}
By induction on the derivation length. If $n=0$ then $t=s$ and we have $\taille{t} \leq \taille{t}$. Now suppose that it holds for a derivation of length $n-1$ by induction hypothesis, that is $t \to_\Regles^{n-1} s$ and $\taille{s} \leq  \taille{t} \times  \taille{\prog}^{n-1}$ and consider one more rewrite step $s \to_\Regles^{1} u$. By definition of a rewrite step, there is a one-hole context $\conta{\magique}$, a rule $l \to r \in \Regles$ and a substitution $\sigma$ such that $s=\conta{l\sigma} \to_\Regles \conta{r\sigma}=u$. As a result, we obtain that:
\begin{align*}
\taille{u}&=\taille{ \conta{r\sigma}}=\taille{\conta{\magique}}+ \taille{r\sigma} &&\\
& \leq \taille{\conta{\magique}}+ \taille{\prog} \times \emph{\max}_{\varone \in \Var(r)}\taille{\varone\sigma} &&\text{Since } \taille{r} \leq \taille{\prog}\\ 
 &\leq  \taille{\conta{\magique}}+\taille{\prog} \times \taille{l\sigma}  &&\text{Since } \Var(r) \subseteq \Var(l)\\
 &\leq  \taille{s } \times\taille{\prog} &&\text{Since }\taille{\prog} \geq 1
\end{align*}
Combining both inequalities, we obtain $\taille{u} \leq  \taille{s } \times\taille{\prog} \leq \taille{t} \times  \taille{\prog}^{n-1} \times \taille{\prog}$ and so the result.
\end{proof}
Now we relate runtime complexity to sup-interpretations:
\begin{theorem}\label{duconlajoie}
Given a terminating TRS $\prog$, then the assignment $\theta$ defined by:
\begin{itemize}
\item  $\theta(\conone)=0$, if $\conone \in \Cns$ is of arity $0$
\item $\theta(\conone)(\many{X}{n})=\sum_{i=1}^n X_i+1$, if $\conone \in \Cns$ is of arity $n>0$
\item $\theta(\funone)(\many{X}{n})=(\sum_{i=1}^{n} X_i +1 )\times \taille{\prog}^{\mathsf{rc}_\Regles(\sum_{i=1}^{n} X_i +1  )} $, if $\funone \in \Fct$
\end{itemize}
is a sup-interpretation.
\end{theorem}
\begin{proof}
Note that the assignment defined is clearly additive and monotonic. Consequently, we have to show that given a symbol $\funone \in \Fct$ of arity $n$ and values $\many{v}{n} \in \Consterms$, if $\funone(\many{v}{n}) \downarrow$ then $\theta(\funone(\many{v}{n})) \geq \theta(\sem{\funone}(\many{v}{n}))$. Note that by definition of $\theta$, $\forall v \in \Consterms,\ \theta(v)=\taille{v}$. 
Consider the reduction $\funone(\many{v}{n}) \to_\Regles^*  \sem{\funone}(\many{v}{n})$, we know that there exists $m \in \bN$ such that $\funone(\many{v}{n}) \to_\Regles^m \sem{\funone}(\many{v}{n})$ since the TRS is terminating. Consequently:
\begin{align*}
\theta(\sem{\funone}(\many{v}{n})) &= \taille{\sem{\funone}(\many{v}{n})} \\
&\leq \taille{\funone(\many{v}{n})} \times \taille{\prog}^m &&\text{By Lemma~\ref{tyf}}\\
&\leq (\sum_{i=1}^n \taille{v_i}+1) \times \taille{\prog}^{rc_\Regles(\sum_{i=1}^n \taille{v_i}+1)} &&\text{By Def. \ref{dc}}\\
&\leq \theta(\funone)(\taille{v_1},\ldots,\taille{v_n})=\theta(\funone)(\theta(v_1),\ldots,\theta(v_n))
\end{align*}
and so the conclusion.
\end{proof}

\subsection{Sup-interpretations through termination techniques}
Note that the bound provided in Theorem~\ref{duconlajoie} is exponential in the length of a derivation. In general, it cannot be improved since it relies on the fact that most TRS do not compute terms of size polynomial in the reduction length because they make a strong use of variable duplication. However it can be improved by either syntactically restricting the set of considered TRS, considering for example, linear TRS, TRS that do not replicate their variables, or by semantically restricting the shape of the captured TRS wrt some termination technique fixed in advance. Additive polynomial interpretations described in Section 4 are an example of such a tool. Indeed they only capture programs computing polynomial size values. Note that the restriction lies in the additivity condition and no longer holds if we consider arbitrary interpretations.\\
We subsume the main termination techniques that can be used to infer sup-interpretations in the Figure~\ref{dingo}.

\begin{figure}[!h]
\begin{gather*}
\begin{tabular}{|c|c|c|}
\hline
 Termination technique &  SI upper bound & Synthesis problem  \\[2pt] \hline
\hline 
& &  \\
Polynomial interpretations  & $O(2^{2^n})$ & \text{Undecidable} \\[2pt]
\hline
& &  \\
Additive Polynomial interpretations  & $O(n^k)$, $k \in \bN$ & \text{Undecidable} \\[2pt]
\hline
& &  \\
Linear additive interpretations  & $O(n)$ & $\Ptime$ \\[2pt]
\hline
& &  \\
Restricted Matrix interpretations  & $O(n^k)$, $k \in \bN$   & $\NP$ \\[2pt]
\hline 
& & \\
RPO  & $O({f(n)})$, $f \in \mathsf{MR}$  & $\NP$-$\text{complete}^{\dag}$\\[2pt]
\hline 
& & \\
DP-based methods  & $O(2^{f(n)})$, $f \in \mathcal{C}_{\text{DP}}$  & $\NP$ \\[2pt]
\hline 
\end{tabular}
\end{gather*}
\label{dingo}
\caption{}
\end{figure}
In this Figure, the first column lists the termination tool under consideration. The second column provides an upper bound $O(g)$ on the sup-interpretations functions that can be computed with respect to the termination technique under consideration. More precisely, for a n-ary function symbol $\funone$ of a TRS whose termination has been shown using some fixed technique, it means that $\theta(\funone)(X_1,\ldots,X_n)=h(\emph{\max}_{1\leq i \leq n}(X_i))$ is a sup-interpretation, for some function $h$ such that $h =O(g)$. Finally, the last column corresponds to the complexity of the respective termination problem.\\
Now we briefly explain the results of Figure~\ref{dingo} line-by-line:
\begin{itemize}
\item For polynomial interpretations, the doubly exponential upper bound on the derivation length of a terminating TRS was shown by Hofbauer and Lautemann in~\cite{HL88}. An important point to stress is that the obtained result for SI is finer than the one that could be expected by a naive application of Theorem~\ref{duconlajoie}: indeed the SI upper bound remains doubly exponential (and not a triple exponential). The undecidability result of the synthesis is demonstrated in Section~\ref{S4} and was suggested by Lankford~\cite{Lankford}.  
\item If we consider additive polynomial interpretations, the synthesis remains undecidable however the sup-interpretation codomain is restricted to polynomials because the size of a value in $\Consterms$ is exactly equal to its size. This result is due to Bonfante et al.~\cite{BCMT00}. 
\item As a consequence, restriction to linear functions yields a linear upper bound computable in polynomial time using linear programming.
\item For matrix interpretation techniques, \cite{EWZ08} demonstrates that the runtime complexity is exponentially bounded in the height of a term. As a consequence, we obtain a double exponential upper bound when the height equals the size, by a naive application of Theorem~\ref{duconlajoie}. Note that this general framework can be restricted to $O({n^k}),\ k \in \bN$ using polynomially bounded matrix interpretations of~\cite{W10,NZM10} or context dependent interpretations~\cite{MSW08} together with restrictions on the interpretation of constructor symbols, in the same spirit as additive polynomial interpretations. See also~\cite{MMNWZ11} for a generalization of matrix interpretations. The complexity of the synthesis is in $\NP$ because the algorithm that shows the termination of a TRS with matrix interpretations uses a $\mathsf{SAT}$ solver. 
\item The RPO termination technique gives an upper bound exponential in a function $f \in \mathsf{MR}$, where $\mathsf{MR}$ stands for the set of multiple recursive functions. This upper bound relies on the lexicographic comparison which yields a multiple recursive derivation length as demonstrated by~\cite{W95}. This bound can be improved to primitive recursive functions $\mathsf{PR}$ if we restrict to Multiset Path Ordering (MPO) as demonstrated by~\cite{Hof92}. Note that we obtain the required upper bound on SI since both $\mathsf{MR}$ and $\mathsf{PR}$ are closed under exponentiation. The $\NP$-completeness of RPO was demonstrated in~\cite{KN85}. ($\dag$ : Note that contrarily to previous mentioned techniques, this technique shows the existence of a SI but does not provide it explicitly.) \\
\item For DP-based methods, Hirokawa and Moser have demonstrated in~\cite{HM11} that techniques combining Dependency Pairs and restricted Interpretations, named SLI for Strongly Linear Interpretations, yields $O({n^2})$ runtime complexity and, consequently, we obtain sup-interpretations in $O(2^{n^2})$ in this particular case. This technique can be generalized to arbitrarily large upper bounds on SI, depending on the base termination technique used. Consequently, the upper bound is $O(2^{f(n)})$ for some $f \in \mathcal{C}_{\text{DP}}$, where $\mathcal{C}_{\text{DP}}$ is a set of runtime complexity functions induced by the DP-method under consideration. For example, the use of RPO as base technique would give a SI in $ \mathsf{MR}$. Note that in this case, the exponential gap can also be withdrawn by putting extra restrictions on the termination system. The synthesis is also in $\NP$ because the verification is based on $\mathsf{SAT}$ solvers. 
\end{itemize}
One of the main drawbacks in the use of runtime complexity in order to infer sup-interpretations is that we are restricted to terminating TRS and so, to total functions.
From that point of view, it is important to stress that QI and DPI based techniques of Sections~\ref{S5} and~\ref{S6} allow for such a treatment because they do not imply termination even if they are based on polynomial interpretation methods. Consequently, they may allow the programmer to infer (polynomial) space upper bounds on the computed values (and also the intermediate values in the case of QI) even if the derivation length of the considered term is bounded by a function of high complexity.

\section{Conclusion}\label{S8}
In this paper, we have studied three methods (interpretations, quasi-interpre\-tations and DP-interpretations) that define a sup-interpretation. Moreover, we have studied the complexity of the sup-interpretation synthesis problem on sets of polynomials including a max operator and we have shown that some termination techniques may allow the programmer to build sup-interpretations. One important issue that falls outside of the scope of this paper concerns the automation of the synthesis problem: in particular the search of efficient algorithms that could allow the programmer to obtain the sup-interpretation of programs (or TRS computing partial functions). Another important issue is to synthesize sup-interpretations through other techniques (type systems, ...). 
Moreover, we have restricted our discussion to monotonic sup-interpretations. An interesting challenge would be to obtain tighter upper bounds by considering non monotonic functions. It is a very difficult problem since all the techniques known to the author are based on monotonicity conditions. 
Finally, due to lack of space, we have not studied the synthesis problem over other paradigms. However we let the reader check that finding a sup-interpretation can always be turned into a constraint satisfaction, see~\cite{MP08b} for example. Consequently, the complexity results presented in this paper have an impact that is not restricted to term rewriting.

\bibliographystyle{elsarticle-num}
\bibliography{bib}

\end{document}

%% file: macros.tex
\newcommand{\NP}{\mathsf{NP}}
\newcommand{\FPtime}{\mbox{\sc FPtime}}
\newcommand{\FPspace}{\mbox{\sc FPspace}}
\newcommand{\Logspace}{\mbox{\sc Logspace}}
\newcommand{\Ptime}{\mathsf{P}}

\newcommand{\bK}{\mathbb{K}}
\newcommand{\bQ}{\mathbb{Q}^+}
\newcommand{\bR}{\mathbb{R}^+}
\newcommand{\bN}{\mathbb{N}}
\newcommand{\bRp}{\mathbb{R}^+}

\newcommand{\some}[3]{#1_{#2}, \cdots, #1_{#3}}
\newcommand{\many}[2]{\some{#1}{1}{#2}}
\newcommand{\ttsome}[3]{\theta(#1_{#2}), \cdots, \theta(#1_{#3})}
\newcommand{\ttmany}[2]{\ttsome{#1}{1}{#2}}

\newcommand{\p}{\mbox{\textbf{p}}}
\newcommand{\funone}{{\tt f}}
\newcommand{\funtwo}{{\texttt{g}}}
\newcommand{\funon}{\emph{\texttt{f}}}
\newcommand{\funtw}{\emph{\texttt{g}}}

\newcommand{\eone}{{\tt e}}
\edef\termone{e}
\newcommand{\varone}{{\tt x}}
\newcommand{\vartwo}{{\tt y}}

\newcommand{\conone}{{\tt c}}

\newcommand{\cona}{{\tt a}}
\newcommand{\conb}{{\tt b}}
\newcommand{\cond}{{\tt d}}
\def\cont#1#2{\mathsf{#1}[#2]}
\def\conta#1{\cont{C}{#1}}
\def\magique{\diamond}

\newcommand{\Var}{\mathit{Var}}
\newcommand{\Variables}{\Var}
\newcommand{\Fct}{\mathcal{D}}
\newcommand{\Functions}{\Fct}
\newcommand{\Cns}{\mathcal{C}}
\newcommand{\Constructors}{\Cns}
\newcommand{\Regles}{\mathcal{R}}
\newcommand{\Terms}[1]{\mathit{Ter}(#1)}
\newcommand{\Freeterms}{\Terms{\Sigma,\Variables}}
\newcommand{\Freetermsd}{\Terms{\Sigma\cup\{\diamond\},\Variables}}
\newcommand{\Consterms}{\Terms{\Constructors}}
\newcommand{\proga}{\langle \Sigma,
         \Regles  \rangle}
\newcommand{\prog}{\langle \Fct \uplus \Cns,
         \Regles  \rangle}

\newcommand{\MaxPlus}{\mathsf{MaxPlus}}
\newcommand{\MaxPoly}{\mathsf{MaxPoly}}
\newcommand{\kdMaxPoly}{{\mathsf{MaxPoly}}^{(\mathsf{k},\mathsf{d})}}
\newcommand{\kdMaxPlus}{{\mathsf{MaxPlus}}^{(\mathsf{k},\mathsf{d})}} 
\newcommand{\kdmMaxPoly}{{\mathsf{MaxPlus}}^{(\mathsf{k}^{\taille{t}},\mathsf{d}^{\taille{t}})}}
\renewcommand{\max}{\emph{max}}

\newcommand{\taille}[1]{|#1|}
\newcommand{\sem}[1]{\llbracket #1 \rrbracket}
\newcommand{\interp}[1]{\llparenthesis #1 \rrparenthesis}
\newcommand{\inter}[1]{\llceil #1 \rrceil}

\newcommand{\mult}{\texttt{mult}}
\newcommand{\add}{\texttt{add}}
\newcommand{\id}{\texttt{id}}
\newcommand{\sexp}{ \emph{\texttt{exp}}}
\newcommand{\double}{\emph{\texttt{d}}}
\newcommand{\zero}{{\tt 0}}
\newcommand{\true}{\mbox{\textbf{True}}}
\newcommand{\false}{\mbox{\textbf{False}}}
\newcommand{\half}{\emph{\texttt{half}}}
\newcommand{\loga}{\emph{\texttt{log}}}

%% file: main.bbl
\begin{thebibliography}{10}
\expandafter\ifx\csname url\endcsname\relax
  \def\url#1{\texttt{#1}}\fi
\expandafter\ifx\csname urlprefix\endcsname\relax\def\urlprefix{URL }\fi
\expandafter\ifx\csname href\endcsname\relax
  \def\href#1#2{#2} \def\path#1{#1}\fi

\bibitem{MP09}
J.-Y. Marion, R.~P{\'e}choux, {Sup-interpretations, a semantic method for
  static analysis of program resources}, ACM TOCL 10~(4) (2009) 1--31.

\bibitem{Terese}
M.~Bezem, J.~Klop, R.~de~Vrijer, {Term rewriting systems}, Cambridge University
  Press, 2003.

\bibitem{BaaderNipkow98}
F.~Baader, T.~Nipkow, Term Rewriting and All That, Cambridge University Press,
  1998.

\bibitem{Rogers87}
H.~Rogers~Jr, {Theory of recursive functions and effective computability}, MIT
  Press, 1987.

\bibitem{JonesCC}
N.~D. Jones, Computability and complexity, from a programming perspective, MIT
  press, 1997.

\bibitem{AG00}
T.~Arts, J.~Giesl, Termination of term rewriting using dependency pairs, Theor.
  Comput. Sci. 236 (2000) 133--178.

\bibitem{BMMP05}
G.~Bonfante, J.-Y. Marion, J.-Y. Moyen, R.~P\'echoux, Synthesis of
  quasi-interpretations, LCC2005, LICS affiliated Workshop.

\bibitem{Lankford}
D.~Lankford, On proving term rewriting systems are noetherian, Tech. rep.
  (1979).

\bibitem{MN70}
Z.~Manna, S.~Ness, On the termination of {M}arkov algorithms, in: Third hawaii
  international conference on system science, 1970, pp. 789--792.

\bibitem{CL92}
E.~Cichon, P.~Lescanne, Polynomial interpretations and the complexity of
  algorithms, in: CADE, no. 607 in LNAI, 1992, pp. 139--147.

\bibitem{HL88}
D.~Hofbauer, C.~Lautemann, Termination proofs and the length of derivations
  (preliminary version), in: RTA, Vol. 355 of LNCS, Springer, 1989, pp.
  167--177.

\bibitem{MS08}
G.~Moser, A.~Schnabl, {Proving quadratic derivational complexities using
  context dependent interpretations}, in: RTA, Vol. 5117 of LNCS, Springer,
  2008, pp. 276--290.

\bibitem{MM00}
J.-Y. Marion, J.-Y. Moyen, Efficient first order functional program interpreter
  with time bound certifications, in: LPAR, Vol. 1955 of LNCS, Springer, 2000,
  pp. 25--42.

\bibitem{BD10}
G.~Bonfante, F.~Deloup, {Complexity Invariance of Real Interpretations}, in:
  TAMC, Vol. 6108 of LNCS, {S}pringer, 2010, pp. 139--150.

\bibitem{BMM01}
G.~Bonfante, J.-Y. Marion, J.-Y. Moyen, On lexicographic termination ordering
  with space bound certifications, in: PSI, Vol. 2244 of LNCS, Springer, 2001,
  pp. 482--493.

\bibitem{BMM05}
G.~Bonfante, J.-Y. Marion, J.-Y. Moyen, Quasi-interpretations and small space
  bounds., in: RTA, Vol. 3467 of LNCS, Springer, 2005, pp. 150--164.

\bibitem{MP08}
J.-Y. Marion, R.~P{\'e}choux, Characterizations of polynomial complexity
  classes with a better intensionality, in: PPDP, ACM, 2008, pp. 79--88.

\bibitem{Amadio03}
R.~Amadio, Synthesis of max-plus quasi-interpretations, Fundamenta Informaticae
  65~(1--2).

\bibitem{F07}
C.~Fuhs, J.~Giesl, A.~Middeldorp, P.~Schneider-Kamp, R.~Thiemann, H.~Zankl, Sat
  solving for termination analysis with polynomial interpretations, in: SAT,
  LNCS, Springer, 2007, pp. 340--354.

\bibitem{F08}
C.~Fuhs, J.~Giesl, A.~Middeldorp, P.~Schneider-Kamp, R.~Thiemann, H.~Zankl,
  Maximal termination, in: RTA, Vol. 5117 of LNCS, Springer, 2008, pp.
  110--125.

\bibitem{B11}
C.~Borralleras, S.~Lucas, A.~Oliveras, E.~Rodr\'{\i}guez-Carbonell, A.~Rubio,
  Sat modulo linear arithmetic for solving polynomial constraints, J. Autom.
  Reasoning 48~(1) (2012) 107--131.

\bibitem{AMS08}
M.~Avanzini, G.~Moser, A.~Schnabl, Automated implicit computational complexity
  analysis (system description), in: IJCAR, Vol. 5195 of LNCS, 2008, pp.
  132--138.

\bibitem{BMP07}
G.~Bonfante, J.-Y. Marion, R.~P{\'e}choux, Quasi-interpretation synthesis by
  decomposition, in: ICTAC, Vol. 4711 of LNCS, Springer, 2007, pp. 410--424.

\bibitem{Kleene88}
S.~Kleene, {Introduction to metamathematics}, Wolters-Noordhoff, 1988.

\bibitem{D79}
N.~Dershowitz, {A note on simplification orderings}, Information Processing
  Letters 9~(5) (1979) 212--215.

\bibitem{Lucas06}
S.~Lucas, On the relative power of polynomials with real, rational, and integer
  coefficients in proofs of termination of rewriting, Appl. Algebra Eng.
  Commun. Comput. 17~(1) (2006) 49--73.

\bibitem{CMTU05}
E.~Contejean, C.~March{\'e}, A.~Tom{\'a}s, X.~Urbain, Mechanically proving
  termination using polynomial interpretations, J. Autom. Reasoning 34~(4)
  (2005) 325--363.

\bibitem{Lucas07}
S.~Lucas, Practical use of polynomials over the reals in proofs of termination,
  in: PPDP, ACM, 2007, pp. 39--50.

\bibitem{NM10}
F.~Neurauter, A.~Middeldorp, Polynomial interpretations over the reals do not
  subsume polynomial interpretations over the integers, in: RTA, Vol.~6 of
  LIPIcs, Springer, 2010, pp. 243--258.

\bibitem{Matiyasevich93}
Y.~V. Matiyasevich, Hilbert's 10th Problem, Foundations of Computing Series,
  MIT Press, 1993.

\bibitem{Tarski51}
A.~Tarski, A Decision Method for Elementary Algebra and Geometry, University of
  California Press, 1951.

\bibitem{Collins75}
G.~E. Collins, Quantifier elimination for real closed fields by cylindrical
  algebraic decomposition, in: Conference on Automata Theory and Formal
  Languages, Vol.~33 of LNCS, 1975.

\bibitem{HRS90}
J.~Heintz, M.-F. Roy, P.~Solerno, Sur la complexit\'e du principe de
  \texttt{T}arski-\texttt{S}eidenberg, Bulletin de la S.M.F., tome 118 (1990)
  101--126.

\bibitem{BMM07}
G.~Bonfante, J.-Y. Marion, J.-Y. Moyen, Quasi-interpretations a way to control
  resources, Theor. Comput. Sci. 412~(25) (2011) 2776--2796.

\bibitem{M93}
Y.~Matiyasevich, M.~Davis, {Hilbert's tenth problem}, Vol.~94, MIT press, 1993.

\bibitem{W95}
A.~Weiermann, Termination proofs for term rewriting systems by lexicographic
  path orderings imply multiply recursive derivation lengths, Theor. Comput.
  Sci. 139~(1{\&}2) (1995) 355--362.

\bibitem{BCMT00}
G.~Bonfante, A.~Cichon, J.-Y. Marion, H.~Touzet, Algorithms with polynomial
  interpretation termination proof, Journal of Functional Programming 11~(1)
  (2001) 33--53.

\bibitem{EWZ08}
J.~Endrullis, J.~Waldmann, H.~Zantema, Matrix interpretations for proving
  termination of term rewriting, J. Autom. Reasoning 40~(2-3) (2008) 195--220.

\bibitem{W10}
J.~Waldmann, Polynomially bounded matrix interpretations, in: RTA, Vol.~6 of
  LIPIcs, 2010, pp. 357--372.

\bibitem{NZM10}
F.~Neurauter, H.~Zankl, A.~Middeldorp, Revisiting matrix interpretations for
  polynomial derivational complexity of term rewriting, in: LPAR (Yogyakarta),
  Vol. 6397 of LNCS, Springer, 2010, pp. 550--564.

\bibitem{MSW08}
G.~Moser, A.~Schnabl, J.~Waldmann, Complexity analysis of term rewriting based
  on matrix and context dependent interpretations, in: FSTTCS, Vol.~2 of
  LIPIcs, 2008, pp. 304--315.

\bibitem{MMNWZ11}
A.~Middeldorp, G.~Moser, F.~Neurauter, J.~Waldmann, H.~Zankl, Joint spectral
  radius theory for automated complexity analysis of rewrite systems, in: CAI,
  Vol. 6742 of LNCS, Springer, 2011, pp. 1--20.

\bibitem{Hof92}
D.~Hofbauer, Termination proofs with multiset path orderings imply primitive
  recursive derivation lengths, Theor. Comput. Sci. 105~(1) (1992) 129--140.

\bibitem{KN85}
M.~S. Krishnamoorthy, P.~Narendran, On recursive path ordering, Theor. Comput.
  Sci. 40~(2-3) (1985) 323--328.

\bibitem{HM11}
N.~Hirokawa, G.~Moser, Automated complexity analysis based on the dependency
  pair method, in: IJCAR, Vol. 5195 of LNCS, Springer, 2008, pp. 364--379.

\bibitem{MP08b}
J.-Y. Marion, R.~P{\'e}choux, Analyzing the implicit computational complexity
  of object-oriented programs, in: FSTTCS, Vol.~2 of LIPIcs, 2008, pp.
  316--327.

\end{thebibliography}
